\documentclass[twocolumn,amsmath,amssymb,nopacs,superscriptaddress,nofootinbib,prb]{revtex4-1}

\usepackage[usenames,dvips]{color}
\usepackage{graphicx}
\usepackage{xcolor}
\usepackage{dcolumn}
\usepackage[colorlinks=true,citecolor=magenta,linkcolor=blue]{hyperref}
\usepackage{bm}
\usepackage{mathtools}
\usepackage{esint}
\usepackage{epstopdf}
\usepackage{dsfont}
\usepackage[outline]{contour}
\usepackage{nicefrac}

\newcommand{\bea}{\begin{eqnarray}}
\newcommand{\eea}{\end{eqnarray}}
\newcommand{\bt}{\textbf}

\newcommand{\phd}{\phantom{\dag}}
\newcommand{\ph}{\phantom{.}}

\newcommand{\noi}{\noindent}
\newcommand{\no}{\nonumber}

\begin{document}
\def\v#1{{\bf #1}}
\title{Trapping Majorana Zero Modes in Vortices of Magnetic Texture Crystals\\ Coupled to Nodal Superconductors}

\author{Daniel Steffensen}
\affiliation{Niels Bohr Institute, University of Copenhagen, Jagtvej 128, DK-2200 Copenhagen, Denmark}

\author{Brian M. Andersen}
\affiliation{Niels Bohr Institute, University of Copenhagen, Jagtvej 128, DK-2200 Copenhagen, Denmark}

\author{Panagiotis Kotetes}
\email{kotetes@itp.ac.cn}
\affiliation{CAS Key Laboratory of Theoretical Physics, Institute of Theoretical Physics, Chinese Academy of Sciences, Beijing 100190, China}

\vskip 1cm

\begin{abstract}
We propose a mechanism for opening a full bulk energy gap and inducing vortex Majorana zero modes (MZMs) in nodal superconductors (SCs). We show that this becomes possible by coupling the nodal SC of interest to a magnetic texture crystal. The latter consists of superpositions of magnetic textures which repeat periodically in space according to suitable wave vectors that enable spin-flip scattering between all pairs of nodes of the SC, and thus open a full gap in its bulk energy spectrum. In this event, MZMs can be trapped in spin or shift vortices introduced in the magnetic texture crystal. Our approach is generic and applies to nodal SCs of spin-singlet, -triplet, or -mixed type of pairing. Therefore, it promises to find application in a variety of nodal SCs, where the magnetic textures appear either spontaneously due to electron-electron interactions, or are imposed by nanomagnets, or become induced by coupling to a lattice of localized magnetic moments.
\end{abstract}

\maketitle

\section{Introduction}

The experimental study of bound states in superconductors (SCs) has recently witnessed a reheated in\-te\-rest. This came after a series of pioneering theo\-ry pro\-po\-sals which de\-si\-gna\-ted a plethora of pathways to induce non-Abelian anyons in both intrinsic~\cite{ReadGreen,Volovik99,Ivanov,KitaevUnpaired,FeTeSeTopo1,FeTeSeTopo2} and engineered~\cite{FuKane,SauPRL,AliceaPRB,LutchynPRL,OregPRL,Mong,Vaezi,KlinovajaParafermion} topological SCs. The so-called Majorana zero modes~\cite{HasanKane,QiZhang} (MZMs) are so far the most sought-after excitations of this genre, since they constitute the simplest type of non-Abelian anyons. MZMs are charge-neutral, spatially lo\-ca\-li\-zed, pinned to zero ener\-gy, and enjoy a to\-po\-lo\-gi\-cal protection. In addition, they adhere to Ising exchange sta\-ti\-stics, which open perspectives for fault-tolerant quantum com\-pu\-ting~\cite{KitaevTQC,Nayak,AliceaTQC}. The charge neutra\-li\-ty of MZMs brings SCs forward as ideal candidates to look for them, since their quasiparticle excitations arise from hybridized electrons and holes~\cite{Alicea,CarloRev,Leijnse,Franz,SatoAndo,Aguado,LutchynNatRevMat,PawlakRev,PradaRev}. Expe\-ri\-men\-tal fingerprints that can be associated with MZMs have been already captured in a variety of experimental platforms, these including nanowire~\cite{Mourik,MTEarly,Das,MT,Sven,Fabrizio,Attila,Kontos,Moodera}, topological insulator~\cite{Yacoby,Jinfeng,Molenkamp,Giazotto}, magnetic adatom~\cite{Yazdani1,Ruby,Meyer,Yazdani2,Gerbold,Wiesendanger,Cren,GerboldEUphys} and FeTeSe~\cite{JiaXinYin,hongding1,hongding2,Lingyuan,LingyuanZhu} systems. 

It has been theoretically demonstrated that MZMs can be trapped at various types of 0D defects~\cite{ReadGreen,Volovik99,Ivanov,KitaevUnpaired,FeTeSeTopo1,FeTeSeTopo2,FuKane,VolovikBook,TeoKaneHedgehog,Wimmer,TeoKane,Shiozaki,XJLiu}. About three decades ago, it was theoretically shown by Read and Green~\cite{ReadGreen}, and by Volovik~\cite{Volovik99} in pa\-ral\-lel, that a vortex induced in a chiral $p_x+ip_y$ SC traps a single MZM. More recently, Fu and Kane~\cite{FuKane} proposed that a single MZM appears in a vortex of a conventional SC in proximity to the helical surface states of a 3D time-reversal (TR) inva\-riant topological insulator. However, the vortex defects involved, need not to be introduced in the supercon\-duc\-ting order parameter. Indeed, MZMs are also accessible if vortex defects are introduced in the phase of another complex field or the angle of a two-component vector entering the Hamiltonian. For instance, MZMs have already been predicted to emerge in vortices of the complex order pa\-ra\-me\-ter of superfluid~\cite{VolovikBook} and axion-string~\cite{SatoStrings} condensates, as well as in the angle of a two-component spin-orbit coupling (SOC) vector field~\cite{Fujimoto}. Notably, the scenario of a SOC vortex has been recently invoked as a possible me\-cha\-nism to re\-con\-cile the experimental observations of a pair of MZMs in a platform of magnetic adatoms deposited on the surface of a conventional SC~\cite{Cren}.

An additional crucial feature that a MZM platform is required to possess in order for its arising MZMs to be robust and long-lived, is to be characterized by a fully-gapped bulk energy spectrum. In order to meet this stringent requirement, the vast majority of the abovementioned proposals have relied on the presence of a gapful pairing gap, i.e., of the $s$- and $p_x+ip_y$ types. Concomitantly, this constraint has also almost exclusively di\-scou\-ra\-ged the pursuit of vortex MZMs in an equally abundant class of SCs, namely the nodal SCs~\cite{BrydonSchnyder}. Obviously, what hinders nodal SCs from joining the MZM pursuit at full speed, is that the fol\-lo\-wing pressing question has remained so far unanswered, i.e., what is the suitable phy\-si\-cal mechanism that gaps out the nodes of the nodal SC and simultaneously enables non trivial topological phases and vortex MZMs in particular.

\begin{figure}[t!]
\begin{center}
\includegraphics[width=\columnwidth]{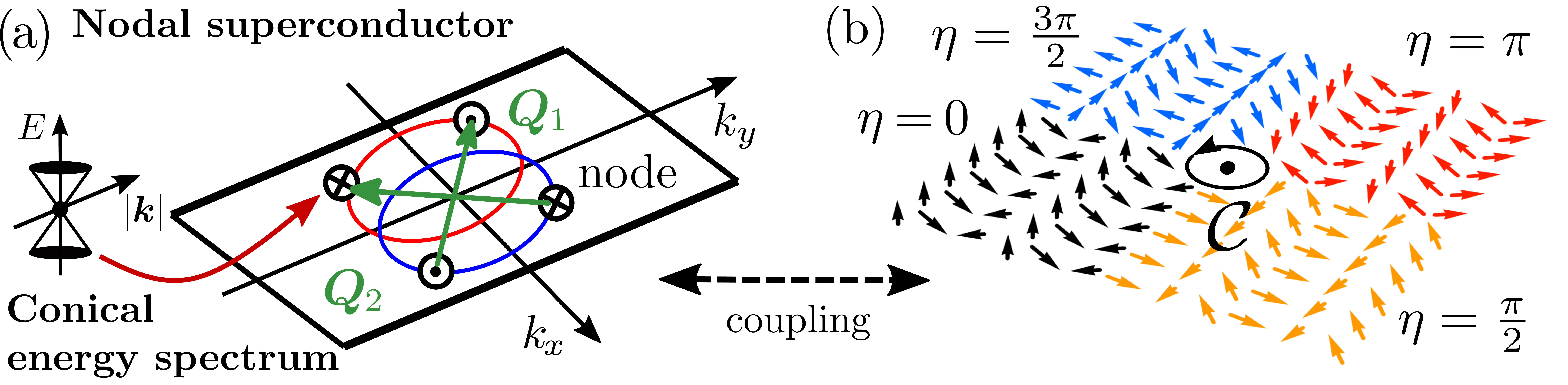}
\end{center}
\caption{(a) Typical bulk energy spectrum for a nodal superconductor discussed here. The nodes come in pairs, and the $n$-th pair is dictated by opposite momenta $\pm\bm{k}_n$ and spin projections $\uparrow,\downarrow$. Thus, the nodes of a given pair carry the same helicity $\zeta=\pm1$ ($\{\bm{\otimes},\bm{\odot}\}$). The nodes are assumed to be subsequently gapped out by the presence of a magnetic helix/stripe texture with a wave vector $\bm{Q}_n$, which may either appear spontaneously due to interactions or be externally imposed. (b) Sketch of a magnetic helix crystal with a spatial profile $\bm{M}\sim\cos[\bm{Q}\cdot\bm{r}+\eta(\bm{r})]\hat{\bm{z}}+\sin[\bm{Q}\cdot\bm{r}+\eta(\bm{r})]\hat{\bm{x}}$. The wave vector is chosen as $\bm{Q}=(2\pi/3,0)$. The texture additionally contains a discrete shift defect with vorticity $\upsilon_{\rm shift}=\sum_{\bm{{\cal C}}}\Delta\eta/2\pi=1$.}
\label{fig:Figure1}
\end{figure}

In this Article, we show that MZMs become accessible in nodal SCs which are under the influence of magnetic texture crystals (MTCs). The MTC is exchange-coupled to the spin of the electrons of the nodal SC and can be either driven by an attractive interaction in the magnetic channel, or, imposed externally to the sy\-stem. MTCs recently got in the spotlight of both theo\-re\-ti\-cal~\cite{KarstenNoSOC,Ivar,KlinovajaGraphene,NadgPerge,KotetesClassi,Nakosai,Braunecker,Klinovaja,Vazifeh,Pientka,Ojanen,Sedlmayr,Mendler,WeiChen,Zutic,Paaske1,Paaske2,Marra,Zutic2,Rex,Zutic3,FPTA} and experimental~\cite{Meyer,PawlakRev,Kontos,Wiesendanger} stu\-dies. Here, we consider MTCs which consist of a superposition of magnetic helices and/or stripes. The $n$-th helix/stripe repeats periodically in space according to a wave vector $\bm{Q}_n$. Each $\bm{Q}_n$ needs to have a length comparable to $2|\bm{k}_n|$, so that it couples and gaps out a pair of nodes of the under\-lying SC at momenta $\pm\bm{k}_n$. In Fig.~\ref{fig:Figure1} we depict a representative nodal bulk energy spectrum where our theory finds application, and we additionally sketch the spatial profile of a MTC containing a shift vortex defect.

Within our proposal, MZMs can be trapped in spin or shift vortices induced in the MTC. In fact, our topological analysis proves that MZMs appear in vortices of the MTC \textit{only} when the underlying SC is nodal. Our theory applies to generic nodal SCs with spin-singlet, -triplet or -mixed~\cite{Gorkov} pai\-ring, thus covering a broad range of quantum ma\-te\-rials and hybrid structures. Remar\-ka\-bly, for systems featuring a Rashba SOC, MZMs can be even trapped by inducing vortices in MTCs which are as mundane as magnetic stripes.

Our upcoming analysis provides a detailed study of the va\-rious topological scenarios that become possible in both 2D and 3D nodal SCs, for all the Majorana symmetry classes, i.e., BDI, D, and DIII. This is achieved by first constructing the topological invariant quantities which predict the emergence of Majorana quasiparticles in two fundamental situations, with these concerning sy\-stems belonging to class BDI in 2D and to class D in 3D. Notably, as we discuss here, nodal SCs coupled to MTCs which belong to class D harbor chiral vortex Majorana modes in 3D. Even more remarkably, we demonstrate that, despite the fact that MTCs break the standard TR symmetry (${\cal T}$), symmetry class DIII SCs and Majorana Kramers pair solutions are still accessible in both 2D and 3D when a generalized TR symmetry $\Theta$ with $\Theta^2=-\mathds{1}$ appears~\cite{KotetesClassi}. Such a symmetry emerges, for instance, when we consider two-band systems, with the electrons of the two bands feeling identical nonmagnetic terms, but opposite MTC terms.

The remainder of this manuscript is organized as follows. Section~\ref{sec:SectionII} contains the details of our theoretical model and sets the stage for our upcoming analysis. Section~\ref{sec:SectionIII} gives an account of the various types of to\-po\-lo\-gi\-cal phases and Majorana excitations which become accessible. In Sec.~\ref{sec:SectionIV} we derive a low-energy model which describes the physics stemming from the nodes of the SC. Based on this low-energy mode, we proceed in Sec.~\ref{sec:SectionV} with the construction of the topological invariant for a class BDI system in 2D, which predicts the emergence of multiple vortex MZMs protected by chiral symmetry. Section~\ref{sec:SectionVI} presents a series of numerical investigations of BDI, D, and DIII models in 2D. In Sec.~\ref{sec:SectionVII} we focus on 3D systems and, in particular, we construct the to\-po\-lo\-gi\-cal invariant for the class D case. Section~\ref{sec:SectionVIII} present numerical results for the emergence of chiral vortex Majorana modes. Section~\ref{sec:SectionIX} gives an account of possible routes to experimentally realize our proposal, while Sec.~\ref{sec:SectionX} concludes this work with a summary and outlook.

\section{Model Hamiltonian}\label{sec:SectionII}

To model the physical situations of interest in a general manner, we employ the Hamiltonian operator:
\begin{align}
\hat{H}=\frac{1}{2}\int d\bm{r}\ph\bm{\Psi}^{\dag}(\bm{r})\hat{\cal H}(\hat{\bm{p}},\bm{r})\bm{\Psi}(\bm{r})\,,
\end{align}

\noi which acts in the basis defined by the Nambu spinor: 
\begin{align}
\bm{\Psi}^{\dag}(\bm{r})=(\psi_\uparrow^{\dag}(\bm{r}),\psi_\downarrow^{\dag}(\bm{r}),\psi_\downarrow(\bm{r}),-\psi_\uparrow(\bm{r}))\,.\label{eq:Spinor}
\end{align}

\noi Here, $\psi_{\uparrow,\downarrow}(\bm{r})$ annihilates an electron at position $\bm{r}$ with the spin projection indicated, while $\hat{\bm{p}}=-i\bm{\nabla}$ with $\hbar=1$. In 3D coordinate space, we define $\bm{r}=(x,y,z)$, $\tan\phi=y/x$, $\cos\theta=z/r$, $r=\sqrt{\rho^2+z^2}$ and $\rho=\sqrt{x^2+y^2}$. 

The matrix $\hat{\cal H}(\hat{\bm{p}},\bm{r})$ defines the Bo\-go\-liu\-bov - de Gennes (BdG) Hamiltonian operator:
\bea
\hat{\cal H}(\hat{\bm{p}},\bm{r})&=&\hat{\cal H}_0(\hat{\bm{p}})+\sum_n\Big\{2M_n\cos\big[\bm{Q}_n\cdot\bm{r}+\eta_n(\bm{r})\big]\hat{\bm{e}}_n\cdot\bm{\sigma}\no\\
&-&2M_n'\sin\big[\bm{Q}_n\cdot\bm{r}+\eta_n(\bm{r})\big]\hat{\bm{e}}_n'\cdot\bm{\sigma}\Big\}e^{-i\omega_n(\bm{r})\sigma_z}\,,\ph\quad\label{eq:FullBdG}
\eea

\noi and is represented using the $\bm{\tau}$ ($\bm{\sigma}$) Pauli matrices defined in Nambu (spin) spaces, supplemented with the respective unit matrix $\mathds{1}_{\tau}$ ($\mathds{1}_{\sigma}$). For simplicity, we omit writing unit matrices throughout from now on. In the above, we consider that the two contributing magnetization terms always feature orthogonal orientations in spin space, i.e., their orientation vectors satisfy $\hat{\bm{e}}_n\cdot\hat{\bm{e}}_n'=0$ for all $n$. 

The nonmagnetic part of the BdG Hamiltonian takes the general form:
\bea
\hat{\cal H}_0(\hat{\bm{p}})
&=&\tau_z\big[\varepsilon_s(\hat{\bm{p}})+\varepsilon_t(\hat{\bm{p}})\sigma_z\big]+\tau_x\big[\Delta_s(\hat{\bm{p}})+\Delta_t(\hat{\bm{p}})\sigma_z\big]\no\\
&+&\tau_y\big[\Delta_s'(\hat{\bm{p}})+\Delta_t'(\hat{\bm{p}})\sigma_z\big]\,,\label{eq:H0}
\eea

\noi where the appearing terms satisfy the relations: 
\bea
\varepsilon_{s,t}(-\hat{\bm{p}})&=&\pm\varepsilon_{s,t}(\hat{\bm{p}})\,,\\ 
\Delta_{s,t}(-\hat{\bm{p}})&=&\pm\Delta_{s,t}(\hat{\bm{p}})\,,\\
\Delta_{s,t}'(-\hat{\bm{p}})&=&\pm\Delta_{s,t}'
(\hat{\bm{p}})\,.
\eea

\noi The above properties imply that $\hat{\cal H}_0(\hat{\bm{p}})$ is invariant under translations and $z$-axis spin rotations, associated with the phases $\eta_n(\bm{r})$ and angles $\omega_n(\bm{r})$, respectively. Vortices can be independently introduced in all $\omega_n$ angles and $\eta_n$ phases, at the same or different positions. 

For a shift [spin] vortex defect with vor\-ti\-ci\-ty $\upsilon_{\rm shift}$ [$\upsilon_{\rm spin}$] we set $\eta(\bm{r})=\upsilon_{\rm shift}\phi$ [$\omega(\bm{r})=\upsilon_{\rm spin}\phi$]. In Fig.~\ref{fig:Figure1}(b) we depict the spatial profile of a magnetic helix crystal with a discrete shift vortex. A shift vortex defect in $\eta(\bm{r})$ implies that this phase shows dis\-con\-ti\-nuous jumps by an integer multiple of $2\pi$ after traver\-sing a closed path ${\cal C}$ encircling the defect's core, which is identified with the region where the magnetic texture va\-ni\-shes. A si\-mi\-lar be\-ha\-vior emerges for $\omega(\bm{r})$ in the presence of a spin vortex. The above properties are reflected in the definitions of the shift $\upsilon_{\rm shift}$ and spin $\upsilon_{\rm spin}$ vorticities:
\bea
\upsilon_{\rm shift}=\ointctrclockwise_{\cal C}\frac{d\eta}{2\pi}\in\mathbb{Z}\quad{\rm and}\quad \upsilon_{\rm spin}=\ointctrclockwise_{\cal C}\frac{d\omega}{2\pi}\in\mathbb{Z}\,.
\eea

\section{Accessible Topological Phases}\label{sec:SectionIII}

To infer the emergence of Majorana quasiparticles in our model, we employ standard classification me\-thods, cf Refs.~\cite{TeoKane,Shiozaki}. The topological classification of the system in the pre\-sen\-ce of defects is carried out using the BdG Hamiltonian in combined momentum-coordinate space $\hat{\cal H}(\bm{k},\bm{r})$, which is obtained by assuming that the defect builds up in a sufficiently smooth manner in space, so that the momentum $\hat{\bm{p}}\mapsto\bm{k}$ and the position $\bm{r}$ appea\-ring in $\eta(\bm{r})$ and $\omega(\bm{r})$ commute. This approach suffices to predict the appea\-ran\-ce of MZMs, but generally fails to accurately describe the complete bound state spectrum that we observe in our numerics using abrupt defects. 

The re\-le\-vant Majorana symmetry class, i.e., BDI, D or DIII, is inferred in the presence of the defect-containing va\-ria\-bles. The effective classification dimension $\delta$ is obtained by the spatial dimensionality of the system $d$, after subtracting the dimension of the surface that can enclose the defect, i.e., here $\delta=d-1$ since a circle $\mathbb{S}^1$ can enclose a vortex. To construct the topological invariants, we view $\phi$ as a synthetic momentum which extends the base space to $(\bm{k},\phi)$.

Based on the tenfold classification
tables~\cite{Altland,KitaevClassi,Ryu}, we find the topologically-nontrivial scenarios $\{{\rm BDI,D,DIII}\}\mapsto\{\mathbb{Z},\mathbb{Z}_2,\mathbb{Z}_2\}$ in 2D, and $\{{\rm D,DIII}\}\mapsto\{\mathbb{Z},\mathbb{Z}_2\}$ in 3D. For the topological description of the cases of relevance in 2D (3D) coordinate space, it suffices to examine the structure of the class BDI (D) $\mathbb{Z}$ topological invariant, which is identified with the winding number $w_3$ (2nd Chern number $C_2$). Here, we obtain general expressions for $w_3$ and $C_2$, which become particularly transparent in the limit of a weak strength for the magnetization of the MTC.

\section{Low-Energy Model Hamiltonian - Derivation}\label{sec:SectionIV}

Similar to Ref.~\cite{XJLiu}, which discusses MZMs trapped in superconducting vortices, also here, the outcome of the various topological invariants is tied to the local, instead of the global, $\bm{k}$-space to\-po\-lo\-gy of $\hat{\cal H}_0(\bm{k})$. Therefore, to facilitate the calculation of the various topological inva\-riants, we rely on low-energy models obtained after expanding the original Hamiltonian about pairs of nodes with momenta $\pm\bm{k}_n$. 

To simplify our upcoming analysis, we momentarily drop the terms $\Delta_{s,p}'(\hat{\bm{p}})$ from the Hamiltonian in Eq.~\eqref{eq:H0}, and restore them for the discussion of 3D models in Sec.~\ref{sec:SectionVIII}. Under the above simplification, the various pairs of nodes are determined by $\hat{\cal H}_0(\bm{k})=\hat{0}$, which boils down to satisfying the condition:
\bea
\varepsilon_s(\bm{k}_n)\pm\sigma_z\varepsilon_t(\bm{k}_n)=\Delta_s(\bm{k}_n)\pm\sigma_z\Delta_t(\bm{k}_n)=\hat{0}\,.
\eea

\noi Since $\{\varepsilon_t(-\bm{k}),\Delta_t(-\bm{k})\}=-\{\varepsilon_t(\bm{k}),\Delta_t(\bm{k})\}$ we find that nodes at opposite momenta $\pm\bm{k}_n$ carry opposite spins $\sigma_z=\pm1$, i.e., possess the same helicity. See Fig.~\ref{fig:Figure1}(a). 

We now expand the Hamiltonian about the $n$-th pair of nodes by setting $\bm{k}\approx\pm\bm{k}_n+\bm{q}$ with $|\bm{q}|\ll|\bm{k}_n|$. By introducing the $\bm{\rho}$ Pauli matrices in $\{\bm{k}_n,-\bm{k}_n\}$ nodes space, the defect-free Hamiltonian in the vicinity of $\pm\bm{k}_n$ reads:
\bea
&&\hat{\cal H}^{(n)}(\bm{q},\phi=0)=M_n\rho_x\hat{\bm{e}}_n\cdot\bm{\sigma}-M_n'\rho_y\hat{\bm{e}}_n'\cdot\bm{\sigma}\no\\
&&+\tau_z\big[\varepsilon_s^{(n)}+\bm{v}_{\varepsilon_t}^{(n)}\cdot\bm{q}\sigma_z\big]
+\rho_z\tau_z\big[\bm{v}_{\varepsilon_s}^{(n)}\cdot\bm{q}+\varepsilon_t^{(n)}\sigma_z\big]\no\\
&&+\tau_x\big[\Delta_s^{(n)}+\bm{v}_{\Delta_t}^{(n)}\cdot\bm{q}\sigma_z\big]+\rho_z\tau_x\big[\bm{v}_{\Delta_s}^{(n)}\cdot\bm{q}+\Delta_t^{(n)}\sigma_z\big],\quad
\label{eq:NodeHamiltonian}
\eea

\noi where we used the shorthand expressions for $f=\varepsilon,\Delta$: 
\bea 
f_{s,t}^{(n)}=f_{s,t}(\bm{k}_n)\phd&{\rm and}&\phd\left.\bm{v}_{f_{s,t}}^{(n)}=\bm{\nabla}_{\bm{k}}f_{s,t}(\bm{k})\right|_{\bm{k}=\bm{k}_n}\,.
\eea

\noi The nonmagnetic part of Eq.~\eqref{eq:NodeHamiltonian}, that we denote $\hat{\cal H}_0^{(n)}(\bm{q})$, is invariant under arbitrary $\phi$-dependent shifts and spin rotations generated by the operators $\hat{\cal L}_{\rm shift}^{(n)}=\rho_z$ and $\hat{\cal L}_{{\rm spin}}^{(n)}=\sigma_z$. Thus, the defects are added as follows:
\bea
\hat{\cal H}^{(n)}(\bm{q},\phi)=e^{i\phi\hat{\cal L}^{(n)}/2}\hat{\cal H}^{(n)}(\bm{q},\phi=0)e^{-i\phi\hat{\cal L}^{(n)}/2}\,,\quad\label{eq:HParametric}
\eea

\noi where we introduced:
\bea
\hat{\cal L}^{(n)}=\upsilon_{\rm shift}^{(n)}\hat{\cal L}_{\rm shift}^{(n)}+\upsilon_{{\rm spin}}^{(n)}\hat{\cal L}_{{\rm spin}}^{(n)}\,.
\eea

For $M_n=M_n'=0$, one defines the four states $\left|\rho_z=\pm1;\sigma_z=\pm1\right>$ in $\rho\otimes\sigma$ space. Two of these give rise to the pair of nodes at $\pm\bm{k}_n$, while the remaining two lie energetically away from zero. These two pairs of states can be distinguished by their he\-li\-ci\-ty eigenvalue $\zeta=\rho_z\sigma_z=\pm1$. Hence, to obtain a Hamiltonian describing only the states related to the nodes, we project Eq.~\eqref{eq:NodeHamiltonian} onto a given helicity subspace which fulfills:
\bea
\varepsilon_s^{(n)}+\zeta\varepsilon_t^{(n)}=\Delta_s^{(n)}+\zeta\Delta_t^{(n)}=0\,,\no
\eea 

\noi and end up with the following effective Hamiltonian for the $n$-th pair of nodes:
\begin{align}
\hat{\cal H}_\zeta^{(n)}(\bm{q},\phi=0)=\lambda_z\bm{q}\cdot\big[\bm{v}_{\varepsilon,\zeta}^{(n)}\tau_z+\bm{v}_{\Delta,\zeta}^{(n)}\tau_x\big]+\bm{M}^{(n)}_\zeta\cdot\bm{\lambda}\,,\label{eq:NodeHamiltonianProjected}
\end{align}

\noi where we introduced the velocities:
\begin{align}
\bm{v}_{\varepsilon,\zeta}^{(n)}=\zeta\bm{v}_{\varepsilon_s}^{(n)}+\bm{v}_{\varepsilon_t}^{(n)}\quad {\rm and}\quad \bm{v}_{\Delta,\zeta}^{(n)}=\zeta\bm{v}_{\Delta_s}^{(n)}+\bm{v}_{\Delta_t}^{(n)} 
\end{align}

\noi along with the parameters:
\begin{align}
\bm{M}_\zeta^{(n)}=\left(\hat{e}_{n,x}M_n+\zeta\hat{e}_{n,y}'M_n',\hat{e}_{n,y}M_n-\zeta\hat{e}_{n,x}'M_n',0\right)\,,
\end{align}

\noi which quantify the influence of the MTC on the given pair of nodes. The unit $\mathds{1}_{\lambda}$ and Pauli $\bm{\lambda}$ matrices act in a given helicity subspace. The choice of basis for both $\zeta=\pm1$ is such, so that the spin Pauli matrix $\sigma_z$ coincides with $\lambda_z$. Note that the terms $\hat{\bm{e}}_n\cdot\hat{\bm{z}}$ and $\hat{\bm{e}}_n'\cdot\hat{\bm{z}}$ drop out after the projection. Projecting the operator generating the vortices yields:
\begin{align}
\hat{\cal L}_\zeta^{(n)}=\big[\zeta\upsilon_{\rm shift}^{(n)}+\upsilon_{\rm spin}^{(n)}\big]\lambda_z\,.\label{eq:NodeOperatorProjected}
\end{align}

\noi Notably, the emergence of MZMs is guaranteed by the structure of Eqs.~\eqref{eq:NodeHamiltonianProjected} and~\eqref{eq:NodeOperatorProjected}, which allow mapping our model to the Jackiw-Rossi model~\cite{JRossi}. The latter is known to support zero-energy solutions in vortices, and also lies at the core of the Fu-Kane MZM proposal~\cite{FuKane,Chamon}. 

\section{Low-Energy Model Hamiltonian - BDI Class Topological Invariant in 2D}\label{sec:SectionV}

In this paragraph we prove in a detailed fashion that MZMs become accessible in the model of Eq.~\eqref{eq:NodeHamiltonianProjected}. The Hamiltonian in Eq.~\eqref{eq:NodeHamiltonianProjected} possesses a chiral symmetry effected by the operator $\Pi=\lambda_z\tau_y$. As a result of it, the Hamiltonian resides in class BDI and is classified by the winding number~\cite{TeoKane} $w_3^{(n)}\in\mathbb{Z}$ defined in $(q_x,q_y,\phi)$ space. This inva\-riant is calculated using the upper off-diagonal block $\hat{h}_\zeta^{(n)}(\bm{q},\phi)$ of $\hat{\cal H}_\zeta^{(n)}(\bm{q},\phi)$, in a basis where the latter is block off-diagonal. The winding number is defined as:
\bea
w_3&=&\int_0^{2\pi}\frac{d\phi}{2\pi}\int\frac{d\bm{q}}{2\pi}\ph{\rm Tr}
\Big\{\hat{h}_\zeta^{-1}(\bm{q},\phi)\big[\partial_{q_x}\hat{h}_\zeta(\bm{q},\phi)\big]\qquad\no\\
&&\hat{h}_\zeta^{-1}(\bm{q},\phi)\big[\partial_{q_y}\hat{h}_\zeta(\bm{q},\phi)\big]\hat{h}_\zeta^{-1}(\bm{q},\phi)\big[\partial_{\phi}\hat{h}_\zeta(\bm{q},\phi)\big]\Big\}\,,\qquad\label{eq:windingBasic}
\eea

\noi where we momentarily drop the $^{(n)}$ index for simplicity. Using the relation $\hat{h}_\zeta\hat{h}_\zeta^{-1}=\mathds{1}\Rightarrow \partial h_\zeta^{-1}=-\hat{h}_\zeta^{-1}(\partial\hat{h}_\zeta)\hat{h}_\zeta^{-1}$ and the cyclic property of the trace, we find the equivalent expression:
\bea
w_3&=&-\int_0^{2\pi}\frac{d\phi}{2\pi}\int \frac{d\bm{q}}{2\pi}\ph{\rm Tr}\Big\{\big[\partial_{q_x}\hat{h}_\zeta(\bm{q},\phi)\big]\no\\
&&\qquad\qquad\hat{h}_\zeta^{-1}(\bm{q},\phi)\big[\partial_{q_y}\hat{h}_\zeta(\bm{q},\phi)\big]\partial_\phi\hat{h}_\zeta^{-1}(\bm{q},\phi)\Big\}.\qquad
\eea

\noi Since the following relation also holds:
\bea
\hat{h}_\zeta(\bm{q},\phi)=e^{i\phi\hat{\cal L}/2}\hat{h}_\zeta(\bm{q},\phi=0)e^{-i\phi\hat{\cal L}/2}\,,
\eea

\noi the winding number obtains the simplified form:
\bea
&&w_3
=\int\frac{d\bm{q}}{2\pi i}\ph{\rm Tr}\Bigg\{\frac{\hat{\cal L}}{4}
\Big\{\big[\partial_{q_x}\hat{h}_\zeta(\bm{q},\phi=0)\big]\big[\partial_{q_y}\hat{h}_\zeta^{-1}(\bm{q},\phi=0)\big]\no\\
&&\phd\ph\quad-\big[\partial_{q_x}\hat{h}_\zeta^{-1}(\bm{q},\phi=0)\big]\big[\partial_{q_y}\hat{h}_\zeta(\bm{q},\phi=0)\big]
\Big\}-q_x\leftrightarrow q_y\Bigg\}\no\\
&&\ph\quad=\int \frac{d\bm{q}}{2\pi}\ph{\rm Tr}\bigg\{\frac{\hat{\cal L}}{2}{\large\mathrel{\contour{black}{${\circlearrowleft}$}}}_{q_xq_y}i\ln\big[\hat{h}_\zeta(\bm{q},\phi=0)\big]\bigg\}\,,
\eea

\noi where we introduced the shorthand notation:
\begin{align}
{\large\contour{black}{$\circlearrowleft$}}_{q_xq_y}=\partial_{q_x}\partial_{q_y}-\partial_{q_y}\partial_{q_x}\,,\no
\end{align}

\noi for the dif\-fe\-ren\-tial operator defining vor\-ti\-ci\-ty in $\bm{q}$ space.

The above expression is nonzero even in the limit of a vanishing strength for the MTC, in which case, $\hat{h}(\bm{q},\phi=0)\mapsto\hat{h}_0(\bm{q})$. When $[\hat{\cal L},\hat{h}_0(\bm{q})]=\hat{0}$, we evaluate the trace by introducing the eigenstates of $\hat{\cal L}$, in which basis, $\hat{h}_0(\bm{q})$ is block diagonal. Hence, by further making use of:
\bea
\hat{h}_\zeta^{(n)}(\bm{q},\phi)=e^{i\phi\hat{\cal L}_\zeta^{(n)}/2}\hat{h}_\zeta^{(n)}(\bm{q},\phi=0)e^{-i\phi\hat{\cal L}_\zeta^{(n)}/2},
\eea

\noi and taking into account that the upper off-diagonal block $\hat{h}_{0;\zeta}^{(n)}(\bm{q})$ of $\hat{\cal H}_{0;\zeta}^{(n)}(\bm{q})$ commutes with  $\hat{\cal L}^{(n)}$, we obtain:
\begin{align}
w_{3;\zeta}^{(n)}=\sum_{\lambda=\pm1}\frac{\zeta\upsilon_{\rm shift}^{(n)}+\upsilon_{{\rm spin}}^{(n)}}{2}\lambda
\int\frac{d\bm{q}}{2\pi}{\large\mathrel{\contour{black}{${\circlearrowleft}$}}}_{q_xq_y}i{\rm tr}\ln\big[\hat{h}_{0;\zeta,\lambda}^{(n)}(\bm{q})\big],
\label{eq:windingSimple}
\end{align}

\noi where we employed the eigenstates $\left|\lambda\right>$ of $\hat{\cal L}_\zeta^{(n)}$, which here coincide with the eigenstates of $\lambda_z=\pm1$. In addition, we accordingly restricted the trace ${\rm Tr}$, to a trace ${\rm tr}$ over the remaining degrees of freedom. By now making use of the identity ${\rm tr}\ln\big[\hat{h}_{0;\zeta,\lambda}^{(n)}(\bm{q})\big]=\ln\det\big[\hat{h}_{0;\zeta,\lambda}^{(n)}(\bm{q})\big]$, we write:
\bea
\det\big[\hat{h}_{0;\zeta,\lambda}^{(n)}(\bm{q})\big]=|\det\big[\hat{h}_{0;\zeta,\lambda}^{(n)}(\bm{q})\big]|e^{-i\varphi_{\zeta,\lambda}^{(n)}(\bm{q})}\,.
\eea

\noi The above implies that Eq.~\eqref{eq:windingSimple} is nonzero only when the arguments $\varphi_{\zeta,\lambda}^{(n)}(\bm{q})$ contain $\bm{q}$-space vortex defects, i.e., \textit{only when the underlying SC contains point nodes.} 

The node with helicity $\zeta$ and $z$-axis spin projection $\sigma_z=\pm1$, carries vor\-ti\-ci\-ty $\upsilon_{\zeta,\lambda=\pm1}^{(n)}$, which is defined through the relation:
\bea
{\large\contour{black}{$\circlearrowleft$}}_{q_xq_y}\varphi_{\zeta,\lambda}^{(n)}(\bm{q})=2\pi\upsilon_{\zeta,\lambda}^{(n)}\delta(\bm{q})\,,\eea

\noi and leads to the expression:
\bea
w_{3;\zeta}^{(n)}=\sum_{\lambda=\pm1}\frac{\zeta\upsilon_{\rm shift}^{(n)}+\upsilon_{{\rm spin}}^{(n)}}{2}\lambda\upsilon_{\zeta,\lambda}^{(n)}\label{eq:winding}\,.
\eea

\noi To evaluate the above, it is required to determine the vorticities of the nodes. For this purpose, we consider the unitary transformation $(\Pi+\tau_z)/\sqrt{2}$ onto the projected Hamiltonians, and obtain the upper off-diagonal blocks:
\begin{align}
\hat{h}_\zeta^{(n)}(\bm{q},\phi=0)=\big[\bm{M}^{(n)}_\zeta\times\hat{\bm{z}}\big]\cdot\bm{\lambda}-\bm{q}\cdot\big[\bm{v}_{\Delta,\zeta}^{(n)}\lambda_z+i\bm{v}_{\varepsilon,\zeta}^{(n)}\big].\label{eq:NodeUpperBlockProjected}
\end{align}

\noi We use the eigenstates of $\lambda_z\mapsto\lambda=\pm1$ and diagonalize $\hat{h}_{0;\zeta}^{(n)}(\bm{q})$ as $h_{0;\zeta,\lambda}^{(n)}(\bm{q})=-\bm{q}\cdot\big[\lambda\bm{v}_{\Delta,\zeta}^{(n)}+i\bm{v}_{\varepsilon,\zeta}^{(n)}\big]$. Therefore, as long as $\bm{v}_{\varepsilon,\zeta}^{(n)}\times\bm{v}_{\Delta,\zeta}^{(n)}\neq\bm{0}$, the vorticities of the nodes at $\bm{q}=\bm{0}$ are opposite and of a single unit, hence, they satisfy $\upsilon_{\zeta,-\lambda}^{(n)}=-\upsilon_{\zeta,\lambda}^{(n)}$ and $|\upsilon_{\zeta,\lambda}^{(n)}|=1$.

Under the above conditions, we obtain \textit{our main result}:
\bea
w_{3;\zeta}^{(n)}={\rm sgn}\left[\upsilon_{\zeta,\lambda=+1}^{(n)}\right]\left[\zeta\upsilon_{\rm shift}^{(n)}+\upsilon_{\rm spin}^{(n)}\right],\label{eq:w3result}
\eea

\noi which implies that both spin and shift vortex defects can independently induce a $\mathbb{Z}$ number of MZMs. Notably, the number of MZMs arising due to the simultaneous emergence of shift and spin vortices at the same position in coordinate space, are obtained by adding (for $\zeta=1$) or subtracting (for $\zeta=-1$) the number of MZMs that would independently arise for each different type of defect.

\section{Numerical Calculations in 2D}\label{sec:SectionVI}

In this section we numerically verify our above predictions for a variety of models in symmetry classes BDI, D, and DIII. Our starting point for all these investigations is the lattice model defined by the following functions:
\bea
&&\varepsilon_s(\bm{k})=-2t(\cos k_x+\cos k_y)-\mu,\quad\varepsilon_t(\bm{k})=\alpha\sin k_y,\no\\
&&\qquad\qquad\ph\Delta_s(\bm{k})=\Delta\quad {\rm and}\quad\Delta_t(\bm{k})=d_z\sin k_y\,.\quad
\label{eq:Model}
\eea

\noi In the absence of magnetism and for a suitable window of pa\-ra\-me\-ters, this model supports a nodal energy spectrum of the form depicted in Fig.~\ref{fig:Figure1}(a). For our upcoming numerical simulations we consider a $40\times40$ square lattice with the lattice constant set to unity. Moreover, all the energy scales are expressed in units of $t$, which from now on is set to unity.

\subsection{Symmetry Class BDI Models}

\begin{figure}[t!]
\begin{center}
\includegraphics[width=1.0\columnwidth]{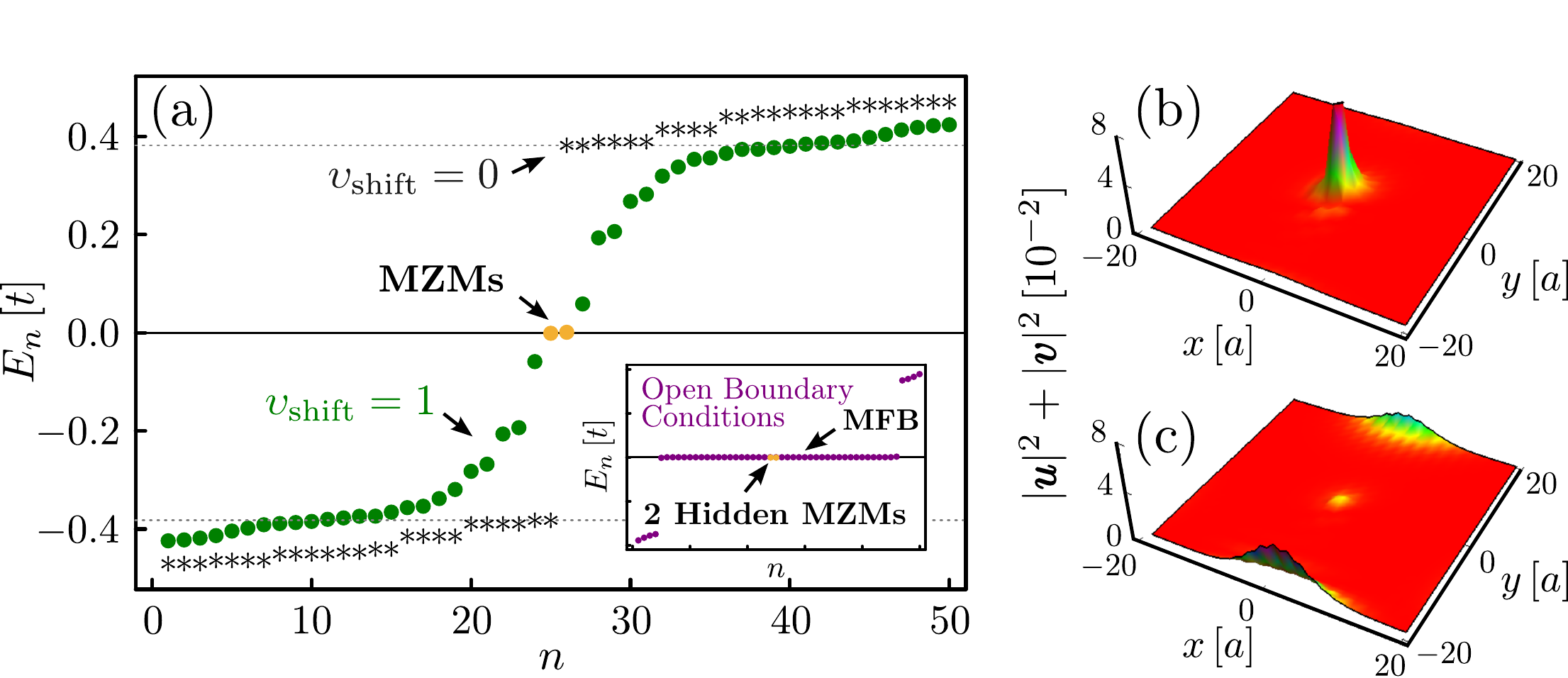}\vspace{0.15in}
\includegraphics[width=1.0\columnwidth]{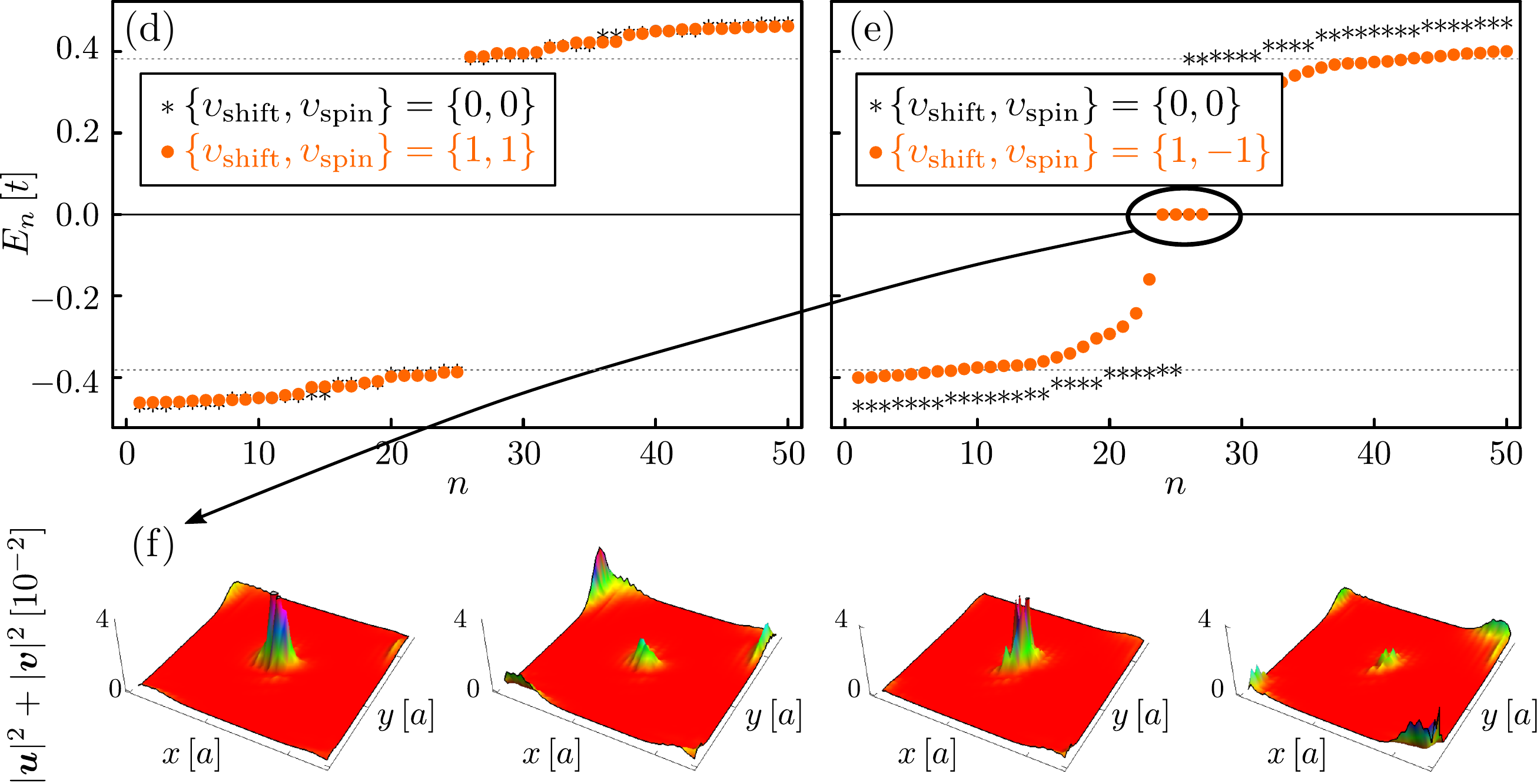}
\end{center}
\caption{(a) 50 lowest eigenvalues in the absence (black asterisks) and presence (green dots) of a single shift vortex in the MTC with $\upsilon_{\rm shift}=1$. When con\-si\-de\-ring open boun\-dary conditions, see inset in (a), we find a single MZM pair along with an edge Majorana flat band (MFB). To uncover the MZMs which are energetically buried inside the MFB, we employ instead periodic boundary conditions. (d)-(e) The $50$ lowest eigenvalues in the absence (black asterisks) and pre\-sen\-ce (orange dots) of a composite spin-shift vortex defect with $\{\upsilon_{\rm shift},\upsilon_{\rm spin}\}=\{1,\pm1\}$, respectively. In accordance to Eq.~\eqref{eq:w3result} only the latter leads to MZMs. (b)-(c) and (f) depict the spatial distribution of the MZM eigenvectors. For extrac\-ting the above numerical results, we considered the following values for the parameters of the model in Eq.~\eqref{eq:Model}: $\Delta=1/\sqrt{2}$, $\mu=-5\Delta$, $d_z=\alpha=1$ and $\{M_{1,2},M_{1,2}'\}=\{0.5,0.1\}$.}
\label{fig:Figure2}
\end{figure}

We begin with the study of MZMs in a BDI class model in 2D. We consider that the two pairs of nodes emerging in the energy spectrum of the model in Eq.~\eqref{eq:Model} get gapped out by a MTC which consists of two helices $\bm{M}_{1,2}(\bm{r})$, with wave vectors $\bm{Q}_{1,2}$. In Fig.~\ref{fig:Figure2}, we present results for helices with $\{\hat{\bm{e}}_{1,2},\hat{\bm{e}}_{1,2}'\}=\{\hat{\bm{x}},\hat{\bm{y}}\}$, when only one of these two magnetic helices harbors a shift vortex defect of a single unit of vorticity.

In accordance with the analytical predictions of Eq.~\eqref{eq:w3result}, our numerical results presented in Fig.~\ref{fig:Figure2}(a)-(c) confirm the emergence of a single MZM pair. One of the MZMs is trapped at the shift defect's core, while the other appears at the system's edge. Moreover, the MZM pair comes along with an edge Majorana flat band (MFB). The latter results from the nodal character of the SC and would anyhow be present at the edge of the system independently of the presence of the vortex.

We remark that the vortex MZM and the MFB do not couple since their wave functions have negligible spatial overlap. Panels (b) and (c) of Fig.~\ref{fig:Figure2} depict the spatial distribution of the eigenvectors for the two MZMs found in (a). We denote the electron (hole) co\-lumn component of the eigenvectors with $\bm{u}$ ($\bm{v}$). One observes that finite-size effects introduced a weak inter-MZM coupling, which in turn leads to a small but nonzero MZM weight at the defect in (c).

\begin{figure*}[t!]
\centering
\includegraphics[width=1\textwidth]{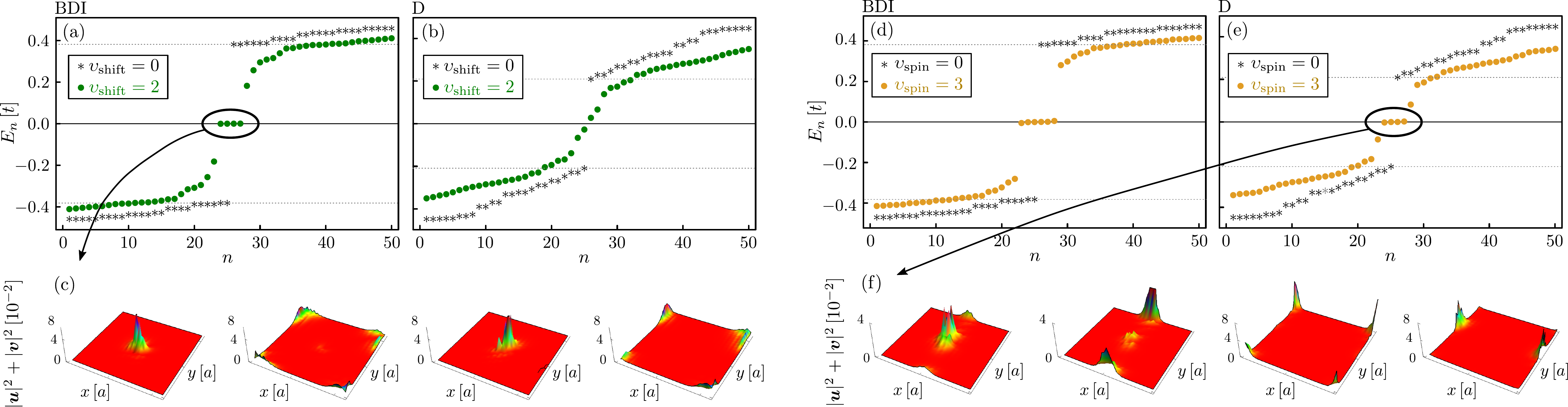}
\caption{(a)-(b) The $50$ lowest eigenvalues in the absence (black asterisks) and presence (green dots) of a shift vortex defect with a two units of vorticity $\upsilon_{\rm shift}=2$ for a class BDI and D model, respectively. In (b) we observe that a class D model does not support MZMs for $\upsilon_{\rm shift}=2$, in agreement with the invariant defined as the parity of $w_{3;\zeta}^{(n)}$. As indicated by the arrow, we show in (c) the weight of the MZM wavefunctions, where we clearly see two states located at the defect and their charge-conjugate counterparts at the edge of the system. (d)-(f) Same as in (a)-(c) but in the case of a single spin defect with three units of vorticity $\upsilon_{\rm spin}=3$. For the class D model in (e) we expect a single MZM pair, in agreement with the parity of $w_{3;\zeta}^{(n)}$, yet we observe four zero energy states. The additional two states, see the two last panels in (f), are an artifact of the phase jump at the edges of the system, and are therefore not located at the defect. For our numerics we employed the same parameter values as in Fig.~\ref{fig:Figure2}. For the class D cases, we considered an additional magnetic field of strength $B_z=0.4$.}
\label{fig:Figure3}
\end{figure*}

Furthermore, in Fig.~\ref{fig:Figure2}(d)-(f) we numerically confirm the predictions of Eq.~\eqref{eq:w3result} in the case of simultaneous spin and shift vortex defects appearing at the same location in real coordinate space $\bm{r}$. According to Eq.~\eqref{eq:w3result}, the final outcome for the winding number $w_{3;\zeta}$ for the pair of nodes experiencing the combined vortex defects is given by adding or subtracting the independent contributions of each vortex to the winding number. Whether these add up or subtract solely depends on which helicity eigenvalue $\zeta=+1$ or for $\zeta=-1$ characterizes the nodes of interest. Specifically, for the model of Eq.~\eqref{eq:Model} and the parameter values employed, we conclude that $\zeta=-1$. This is directly inferrable from the energy spectra shown in Figs.~\ref{fig:Figure2}(d) and~(e) for $\{\upsilon_{\rm shift},\,\upsilon_{\rm spin}\}=\{1,\,\pm1\}$, where we find that only the latter composite vortex configuration leads to MZMs. In fact, one obtains a total of four MZMs, with two of these being located at the center of the defect, and two more near the edge. This becomes further tran\-spa\-rent from Fig.~\ref{fig:Figure2}(f) where we depict the spatial weights of the MZM eigenvectors. We observe once again that the four MZMs appearing in Fig.~\ref{fig:Figure2}(e) are weakly coupled due to finite-size effects.

The two MZMs comprising the MZM pair appea\-ring at the core of the composite spin-shift vortex with $\{\upsilon_{\rm shift},\,\upsilon_{\rm spin}\}=\{1,\,-1\}$ do not couple to each other by virtue of the chiral symmetry which dictates the system. Notably, multiple uncoupled MZMs trapped at the core of the vortex are also expected to appear for a single shift/spin vortex when this carries a higher value of vorticity. To further verify this prediction, we study va\-rious cases numerically, by implementing the same lattice Hamiltonian defined by Eq.~\eqref{eq:Model}. In Figs.~\ref{fig:Figure3}(a) and (c) we confirm that $\upsilon_{\rm shift}=2$ results into two pairs of MZMs, with two at the center of the defect, and their two partners on the edge. Additionally we confirm that spin defects also lead to MZMs. See Fig.~\ref{fig:Figure3}(d) with 3 pairs of MZMs now appearing at the vortex core for $\upsilon_{\rm spin}=3$.

\subsection{Class D models in 2D} 

The result of Eq.~\eqref{eq:w3result} obtained earlier is valid \textit{only} as long as also the full Hamiltonian resides in class BDI. In fact, it is straigthforward to verify that the full Hamiltonian possesses a chiral symmetry effected by $\tau_y\sigma_z$, only when $\bm{e}_n$ and $\bm{e}_n'$ lie in the same spin plane for all $n$. When at least one of $\varepsilon_t(\bm{k})$ or $\Delta_t(\bm{k})$ is present, this is identified with the $xy$ spin plane. As long as the above condition is met, Eq.~\eqref{eq:w3result} remains valid. 

For a full Hamiltonian belon\-ging to class D instead, solely the pa\-ri\-ty of the winding number $(-1)^{w_{3;\zeta}^{(n)}}\in\mathbb{Z}_2$ is well defined, and allows for only up to a single MZM to be trapped at the core of a vortex defect. 

The above conclusions further imply that particular caution needs to be paid on the possible node degeneracies which can tri\-via\-li\-ze the $\mathbb{Z}_2$ invariant. This takes place, for instance, when only $\varepsilon_s(\bm{k})$ and $\Delta_s(\bm{k})$ enter $\hat{\cal H}_0(\bm{k})$. In this case, both helicities contribute, i.e., $w_3^{(n)}=\sum_{\zeta=\pm1}w_{3;\zeta}^{(n)}$. This case is trivial in class D, since we find $|w_3^{(n)}|=2|\upsilon_{\rm spin}^{(n)}|$, while in class BDI it predicts spin-degenerate MZM pairs only for spin vortices, as a consequence of the spin-singlet character of the pairing. Analogous results with $|w_3^{(n)}|=2|\upsilon_{\rm shift}^{(n)}|$ are obtained when only $\varepsilon_s(\bm{k})$ and $\Delta_t(\bm{k})$ are considered.

We now proceed with scenarios where a symmetry class transition BDI $\mapsto$ D takes place by explicitly brea\-king the chiral symmetry of Eq.~\eqref{eq:FullBdG}, which is effected by $\tau_y\sigma_z$. In the simplest case, this can be achieved by either applying a magnetic field in the $z$ direction, or, by considering spin-orientation vectors $\bm{e}_n$ and $\bm{e}_{n}'$ which lie in different spin planes. By virtue of the symmetry class reduction, it is only the parity $(-1)^{w_{3;\zeta}^{(n)}}$ which can protect MZMs. 

This is confirmed in Fig.~\ref{fig:Figure3}(b) where we display the ener\-gy spectrum for $\upsilon_{\rm shift}=2$ in the presence of a magnetic field in the $z$ direction, here denoted $B_z$, which enters in the Hamiltonian of Eq.~\eqref{eq:FullBdG} through the term $B_z\sigma_z$. For this case $(-1)^{w_{3;\zeta}^{(n)}}=1$, ultimately resul\-ting into the hybridization of the MZMs, thus lifting them away from zero energy. In stark contrast, if we have an odd number of MZMs, i.e. $(-1)^{w_{3;\zeta}^{(n)}}=-1$, a single pair of MZM persists in the presence of a magnetic field in $z$ direction, as seen in Figs.~\ref{fig:Figure3}(e)-(f). We wish to cla\-ri\-fy that despite the fact that in Fig.~\ref{fig:Figure3}(e) we find four in-gap states, only a single pair corresponds to to\-po\-lo\-gi\-cal\-ly protected MZMs, with only one of these MZMs having its wavefunction weight localized at the defect, as one confirms from Fig.~\ref{fig:Figure3}(f). 

\subsection{Class DIII Models in 2D}

Despite the fact that MTCs break the standard time reversal (${\cal T}$) symmetry, Majorana Kramers pairs are still accessible when a generalized TR symmetry $\Theta$ with $\Theta^2=-\mathds{1}$ appears instead~\cite{KotetesClassi}. In this event, the Hamiltonian is of the DIII type and is classified by a $\mathbb{Z}_2$ to\-po\-lo\-gi\-cal inva\-riant which now predicts the emergence of a single Majorana Kramers pair in a shift/spin vortex. 

Such a symmetry emerges in the previously examined mo\-dels when we consider, for instance, two bands labelled by a and b. After introducing the $\bm{\kappa}$ Pauli matrices in band space, the MTC terms contributing to the BdG Hamiltonian get promoted to matrices in band space, allowing for intra- and inter-band magnetic scattering terms proportional to $\mathds{1}_{\kappa},\,\kappa_z$ and $\kappa_x$, respectively. In the remainder, we consider solely intraband magnetic scat\-te\-ring, with the magnetic texture term being proportional to $\kappa_z$. Hence, the two bands feel opposite contributions from the MTCs. 

After considering that the two bands are dictated by identical nonmagnetic terms given by Eq.~\eqref{eq:H0}, we end up with the following Hamiltonian for a two-band system:
\bea
&&\hat{\cal H}'(\hat{\bm{p}},\bm{r})=\hat{\cal H}_0(\hat{\bm{p}})+\sum_n\Big\{2M_n\cos\big[\bm{Q}_n\cdot\bm{r}+\eta_n(\bm{r})\big]\hat{\bm{e}}_n\no\\
&&\quad-2M_n'\sin\big[\bm{Q}_n\cdot\bm{r}+\eta_n(\bm{r})\big]\hat{\bm{e}}_n'\Big\}\cdot\kappa_z\bm{\sigma}e^{-i\omega_n(\bm{r})\sigma_z}\,.\label{eq:FullBdG2Band}
\eea

\noi Since the above bands are completely decoupled, they yield pairs of Majorana solutions in the defect's core. Specifically, we find that the pair of MZMs is protected by the time-reversal symmetry $\Theta=\kappa_x{\cal T}$. Notably, the above Hamiltonian possesses an additional U(1) symmetry with generator $\kappa_z$, which enlists the present system in class BDI$\oplus$BDI, instead of DIII. Hence, in order to obtain true Majorana Kramers pairs protected by $\Theta$ it is required to include band mixing terms which are simultaneously inva\-riant under the action of $\Theta$.

\begin{figure}[t!]
\centering
\includegraphics[width=1\columnwidth]{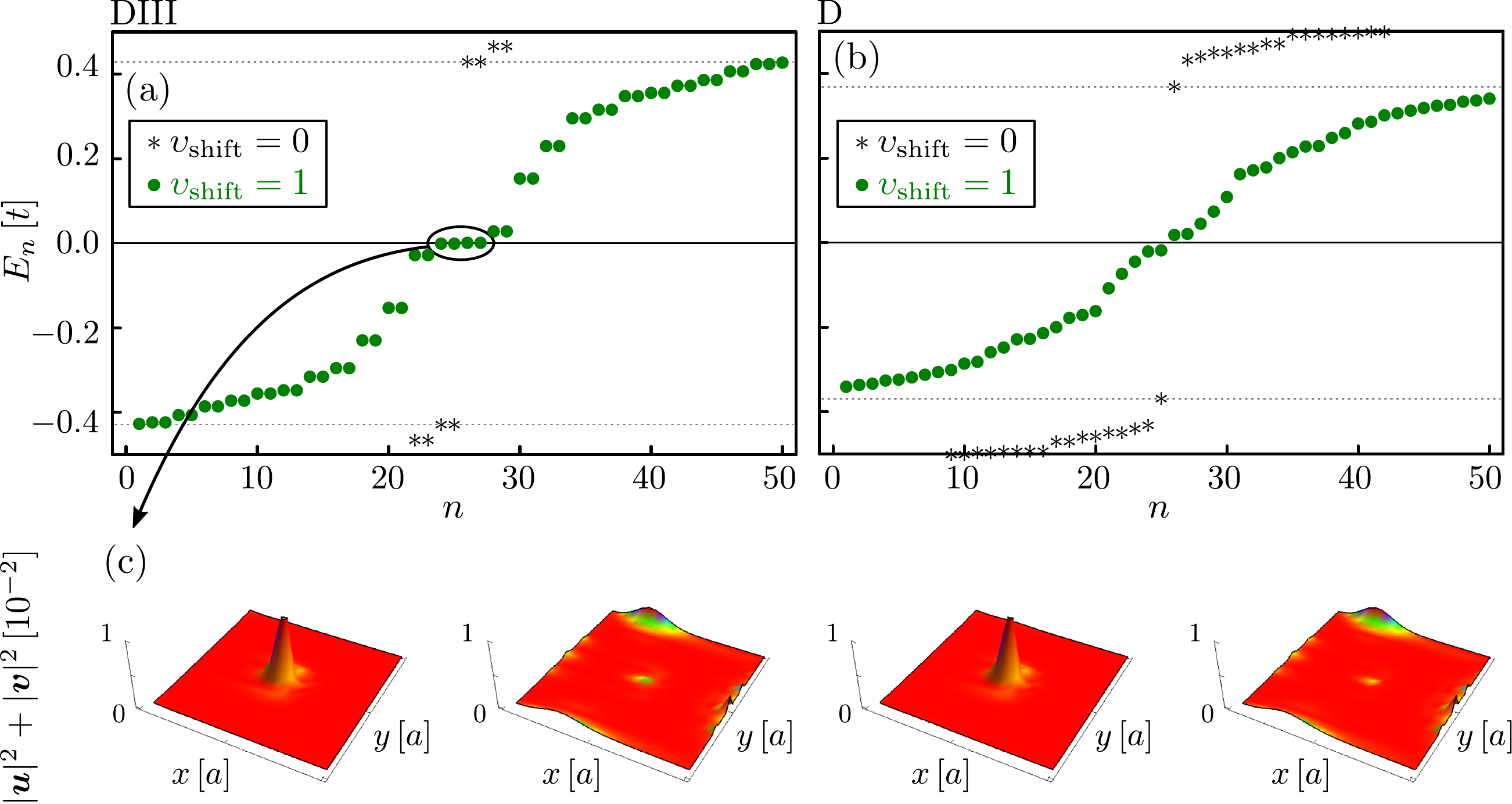}
\caption{Numerical investigation of a 2D class DIII model and its transition to a class D model. (a)-(b) The $50$ lowest eigenvalues in the absence (black asterisks) and presence (green dots) of a shift vortex with $\upsilon_{\rm shift}=1$, for the two-band extension of the BDI model in 2D, ultimately resulting into a class DIII. (c) di\-splays the resulting four zero energy states from the MZM Kramers pair, and (b) shows how these get lifted in the pre\-sen\-ce of a magnetic field of a strength $B_z=0.4$, entering through the term $B_z\sigma_z$. For the DIII model numerics we used: $\mu=-2\sqrt{2}$, $\alpha=2+\mu$, $d_z=1$, $\{M_1,M_1'\}=\{1,0.2\}$, $\{M_2,M_2'\}=\{0.5,0.5\}$, and the band mixing term $\delta_0=0.2\sqrt{2}$. Note that finite-size effects, and inter MZM-coupling result into weights at the defect in the second and fourth panels of (c).}
\label{fig:Figure4}
\end{figure}

In order to numerically study class DIII models in 2D, we consider the lattice extension of the two band Hamiltonian in Eq.~\eqref{eq:FullBdG2Band}, in the additional presence of the weak band mixing term $\delta\tau_z\kappa_x$ which preserves $\Theta$. We focus on the two-band extension of the 2D BDI model in Eq.~\eqref{eq:Model} where, here, we set $\varepsilon_t^{\rm a}(\bm{k})=\varepsilon_t^{\rm b}(\bm{k})=\alpha\sin k_y$, $\Delta_s^{\rm a}(\bm{k})=\Delta_s^{\rm b}(\bm{k})=0$ and $\Delta_t^{\rm a}(\bm{k})=\Delta_t^{\rm b}(\bm{k})=d_z\sin k_x$. Our numerics confirm the emergence of a MZM Kramers pair when considering a single shift/spin vortex defect, as seen in Fig.~\ref{fig:Figure4}(a) where we observe four MZMs. The spatially-resolved MZM wavefunction weights in Fig.~\ref{fig:Figure4}(c) show that one MZM Kramers pair is localized at the defect and another at the outer edge of the system. 

Similarly to the BDI mo\-dels in 2D, we can also here reduce the symmetry of the system by adding a homogeneous external magnetic field. In Fig.~\ref{fig:Figure4}(b) we indeed see that the MZM Kramers pair is lifted away from zero ener\-gy by adding a magnetic field in the $z$ direction, which forces the TR-invariant system to undergo a symmetry-class transition to class D. The latter supports a $\mathbb{Z}_2$ invariant and cannot sustain the MZM Kramers pair.

To this end, we clarify that while $\Theta$ appears to be a fine-tuned symmetry which crucially depends on the here-assumed structure of the MTC, it may actually constitute a robust symmetry in realistic materials. This is for instance the case when the MTC develops spontaneously due to inte\-ractions. In such a scenario, a MTC changing sign on the two bands which is self-generated by the electrons of the SC is expected to be thermodynamically stable, since its emergence leads to a global minimum of the free energy the system.

\section{Low-Energy Model Hamiltonian - D Class Topological Invariant in 3D}\label{sec:SectionVII}

Having examined 2D systems in detail, we now proceed with exploring topological scenarios in 3D. Based on our topological analysis in Sec.~\ref{sec:SectionIII}, 3D allows for two possibilities. Chiral/helical Majorana edge modes disper\-sing along a vortex line arising from a class D/DIII Hamiltonian. The construction of a class DIII Hamiltonian is possible by putting together two class D Hamiltonians following the prescription of the previous section. Since DIII class systems can be understood using an even number of class D copies, in the remainder we focus on the topological aspects of a single class D Hamiltonian.

The emergence of a chiral vortex Majorana mode is now predicted by a second Chern number $C_2$ which is defined in 4D $(q_x,q_y,\phi,q_z)\equiv (p_1,p_2,p_3,p_4)$ space and it is given by the following expression:~\cite{TeoKane}
\begin{align}
C_2=-\int\frac{d^4p}{32\pi^2}\ph\epsilon_{nm\ell s}{\rm Tr}\left[\hat{F}_{nm}\hat{F}_{\ell s}\right]\,,
\end{align}

\begin{figure*}[t!]
\centering
\includegraphics[width=\textwidth]{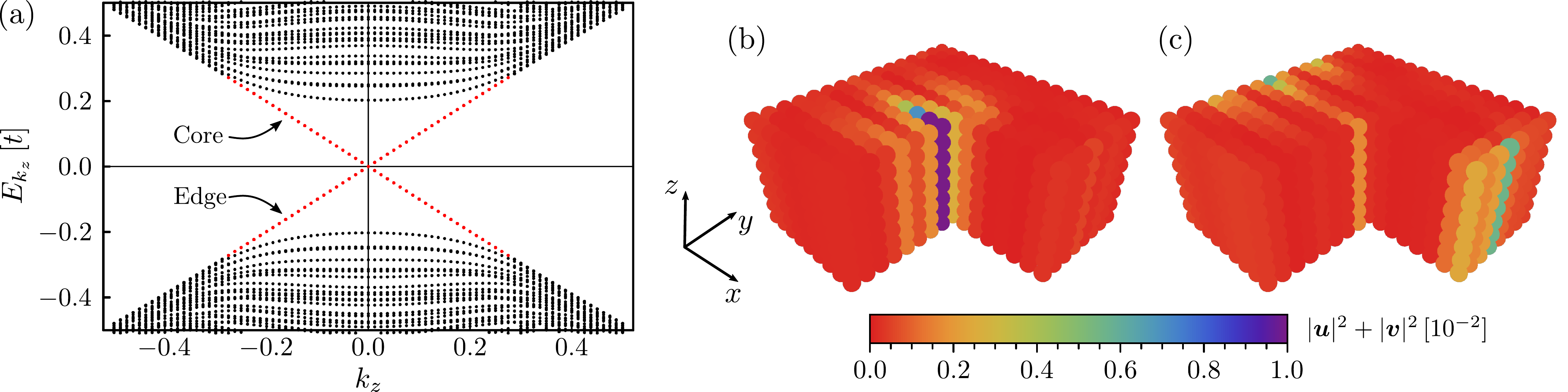}
\caption{Chiral Majorana modes in a class D model in 3D. (a) Edge spectrum for the 3D model with a single shift vortex defect $\upsilon_{\rm shift}=1$. The spectrum is obtained with periodic (open) boundary conditions in the $z$ ($x$ and $y$) direction, and clearly displays chiral Majorana modes. The spatially-resolved weight of the two chiral branches are displayed in (b) and (c), and reveals a single Majorana mode at the vortex defect's core, and its counterpart located at the edge of the system. Note that the nonzero wavefunction weight at the defect in (c), is a consequence of inter-Majorana mode coupling and finite-size effects. The figures were obtained with the parameters: $\Delta=1$, $\mu=-2\sqrt{2}$, $\alpha=2(\sqrt{2}-1)$, $\Lambda=8$ and $\{M_{1,2},M_{1,2}'\}=\{0.5,0.1\}$, all in units of $t$.}
\label{fig:Figure5}
\end{figure*}

\noi with $\epsilon_{nm\ell s}$ denoting the fully-antisymmetric tensor, where $n,m,\ell,s=1,2,3,4$. We also note that repeated index summation was employed. The above formula is given in terms of the non-Abelian field strength tensor:
\begin{align}
\hat{F}_{nm}=\partial_{p_n}\hat{A}_m-\partial_{p_m}\hat{A}_n-i[\hat{A}_n,\hat{A}_m]\,,
\end{align}

\noi which is defined in terms of the Berry vector potential:
\begin{align}
A_n^{\alpha\beta}(\bm{p})=i\left<\bm{\Phi}_\alpha(\bm{p})\right|\partial_{p_n}\left|\bm{\Phi}_\beta(\bm{p})\right>\,,
\end{align}

\noi which is a matrix in the occupied eigenstates $\left|\bm{\Phi}_\alpha(\bm{p})\right>$ subspace, which are enumerated by the index $\alpha$. 

The second Chern number can be equivalently expressed as a surface integral over the Chern-Simons 3 form. Here, we choose a surface ${\cal S}=\mathbb{S}^2\times\mathbb{T}^1$ which contains a $\mathbb{S}^2$ sphere in $\bm{q}$ space. We thus find:
\begin{align}
C_2=-\oint_{\cal S}\frac{d^3p}{8\pi^2}\ph\epsilon_{nm\ell}{\rm Tr}\left(\hat{A}_n\partial_{p_m}\hat{A}_\ell-i\frac{2}{3}\hat{A}_n\hat{A}_m\hat{A}_\ell\right)\,,
\end{align}

\noi with $n,m,\ell=1,2,3$. When the following holds:
\begin{align}
\hat{\cal H}(\bm{q},\phi)=e^{i\phi\hat{\cal L}/2}\hat{\cal H}(\bm{q},\phi=0)e^{-i\phi\hat{\cal L}/2}\,,
\end{align}

\noi we find the relation: $|\bm{\Phi}(\bm{q},\phi)\rangle=e^{i\phi\hat{\cal L}/2}|\bm{\Phi}(\bm{q},\phi=0)\rangle$, which implies $\hat{A}_{\bm{q}}(\bm{q},\phi)=\hat{A}_{\bm{q}}(\bm{q},\phi=0)$ and 
\begin{align}
A_{\phi}^{\alpha\beta}(\bm{q},\phi)=-\frac{1}{2}\left<\bm{\Phi}_\alpha(\bm{q},\phi=0)\right|\hat{\cal L}\left|\bm{\Phi}_\beta(\bm{q},\phi=0)\right>.
\end{align}

\noi The above lead to the simplified expression:
\begin{align}
C_2=\varoiint_{\mathbb{S}^2}\frac{d\bm{q}}{2\pi}\cdot{\rm Tr}\left[\frac{\hat{\cal L}}{2}\hat{\bm{\Omega}}(\bm{q},\phi=0)\right]\,,
\end{align}

\noi where we introduced the matrix Berry curvature $\hat{\bm{\Omega}}(\bm{q},\phi=0)$. The second Chern number is here ge\-ne\-ral\-ly nonzero also for a vanishing MTC strength. Under the assumption $[\hat{\cal L},\hat{\cal H}_0(\bm{q})]=\hat{0}$, we evaluate the trace by introducing the eigenstates of $\hat{\cal L}$, i.e. $\hat{\cal L}\left|\lambda\right>=\upsilon_{\rm defect}\lambda\left|\lambda\right>$, in which basis, $\hat{\cal H}_0(\bm{q})$ and the respective Berry curvature matrix $\hat{\bm{\Omega}}_0(\bm{q})$ of the nonmagnetic system are block diagonal. Thus, we conclude with the expression:
\bea
C_2=\upsilon_{\rm defect}\sum_\lambda\frac{\lambda}{2}\varoiint_{\mathbb{S}^2}\frac{d\bm{q}}{2\pi}\cdot{\rm tr}\left[\hat{\bm{\Omega}}_{0;\lambda}(\bm{q})\right]\,,
\eea

\noi with the trace acting in a given $\lambda$ block. Under the assumption that the SCs under examination possess a zero first Chern number, the second Chern number above becomes nonzero only in the presence of monopoles in the Berry curvature of the SC. These monopoles correspond to $\bm{q}$-space nodes in 3D space, which carry a topological charge defined through ${\rm tr}\big[\hat{\bm{\Omega}}_{0;\lambda}(\bm{q})\big]=Q_\lambda\bm{q}/(2|\bm{q}|^3)$. For a $2\times2$ $\lambda$ block, these monopoles define Weyl points, which carry a topological charge given by:
\bea
\bm{\Omega}_{0;\lambda}(\bm{q})=Q_\lambda\frac{\bm{q}}{2|\bm{q}|^3}\,.
\eea

\noi From the above relation, we obtain the conclusive expression for the invariant, which takes the following form:
\bea
C_{2;\zeta}^{(n)}=\sum_{\lambda=\pm1}\frac{\zeta\upsilon_{\rm shift}^{(n)}+\upsilon_{{\rm spin}}^{(n)}}{2}\lambda Q_{\zeta,\lambda}^{(n)}\,,
\eea

\noi with $Q_{\zeta,\lambda}^{(n)}$ defining the monopole charge for the nodes of the $n$-th pair with he\-li\-ci\-ty $\zeta$ and $z$-axis spin projection $\lambda=\pm1$.

\section{Numerical Calculations for a D Class 3D Model}\label{sec:SectionVIII}
 
The topological invariant $C_2$ obtained in the previous paragraph predicts the number of chiral Majorana modes emerging in the core of a vortex line. We first pursue gaining insight regarding the emergence of such dispersive chiral vortex Majorana modes using the following simple continuum model:
\begin{align}
\hat{\cal H}_0(\bm{k})=-k_y^2\tau_z+\alpha\big(k_y\tau_z+k_x\tau_x-k_z\tau_y\big)\sigma_z\,,
\label{eq:3Dmodel}
\end{align}

\noi which constitutes an anisotropic p-wave SC variant of the model in Eq.~\eqref{eq:Model}. The combination of spatial anisotropy and SOC yields two helical branches and two pair of nodes at $k_y=0$ and $k_y=\pm\alpha$. Here, the inner helical branch at $k_y=0$ can be gapped out by a Zeeman field which is oriented ortho\-go\-nal\-ly to the SOC vector~\cite{LutchynPRL,OregPRL}. The two nodes of the outer helical branch can get gapped out by a magnetic stripe $\bm{M}(\bm{r})=M\cos(2\alpha y)\hat{\bm{x}}$. In ana\-lo\-gy to Eq.~\eqref{eq:w3result}, here we find that a number of:
\begin{align}
|C_2|=|\upsilon_{\rm shift}+\upsilon_{\rm spin}|
\end{align}

\noi chiral Majorana modes emerge in a vortex line extending along the $z$ axis.

We now provide a numerical investigation of the above class D model in 3D, after considering a proper lattice extension. For this purpose, we consider that the bare Hamiltonian of Eq.~\eqref{eq:H0} is given by:
\begin{align}
\hat{{\cal H}}^{\rm 3D}_0(\bm{k})=\tau_z[\varepsilon_s(\bm{k})+\varepsilon_t(\bm{k})\sigma_z]+\big[\Delta_p(\bm{k})\tau_x+\Delta_p'(\bm{k})\tau_y\big]\sigma_z,
\end{align}

\noi and consists of the anisotropic 3D dispersion:
\begin{align}
\varepsilon_s(\bm{k})=-2t(\cos k_x+\cos k_y)-\Lambda(1-\cos k_z)/2-\mu\no
\end{align}

\noi in the additional presence of the anisotropic SOC term $\varepsilon_t(\bm{k})=\alpha\sin k_y$, and the chiral p-wave pairing which is defined in terms of the components:
\bea
\Delta_p(\bm{k})=\Delta\sin k_x\quad{\rm and}\quad \Delta_p'(\bm{k})=-\Delta\sin k_z\,.
\eea

We consider the limit $\Lambda\gg t$, in which, the pairs of nodes in the nonmagnetic phase are located only in the $k_z=0$ plane. After including the magnetic terms of the Hamiltonian and considering a vortex line which extends uniformly along the $z$ axis, we observe that $k_z$ is a good quantum number since $\tan \phi=y/x$. In fact, for small $k_z$, we can linearize the above Hamiltonian and see that for $k_z=0$ it possesses a chiral symmetry with $\Pi=\tau_y\sigma_z$ similar to the model of Eq.~\eqref{eq:Model}. The preservation of $\Pi$ gives rise to a pair of zero ener\-gy states. 

Away from $k_z=0$ the chiral symmetry is broken, lif\-ting the states away from zero ener\-gy, ultimately resulting into dispersive chiral Majorana modes, as seen Fig.~\ref{fig:Figure5}(a). Here, a single mode is disper\-sing along the vortex core while the other one resides on the outer edge of the system, see Fig.~\ref{fig:Figure5}(b) and (c), respectively.

\section{Experimental Implementation}\label{sec:SectionIX}

We now proceed with the discussion of potential candidate systems and mechanisms that may allow obser\-ving our theoretical predictions in realistic systems. As mentioned above, nodal superconducting materials are abundant in nature~\cite{BrydonSchnyder}, e.g., unconventional spin-singlet (-triplet) d-wave (p-wave) SCs~\cite{Sigrist}, noncentrosymmetric~\cite{NCSExpTheory1,NCSExpTheory2,NCSSatoFujimoto,Smidman2017Jan} SCs, and certain Fe-based SCs which can also exhibit nodal pai\-ring~\cite{Clarina,Liu}. Below, we discuss three distinct pathways involving MTCs, which lead to a fully-gapped bulk energy spectrum and open perspectives for vortex MZMs.

\subsection{MTCs Engineered by Nanomagnets}

The first possibility is to actually impose the desired MTC externally with the help of tunable magnets. Such a direction has recently picked up substantial theo\-re\-ti\-cal~\cite{KarstenNoSOC,KlinovajaGraphene,Zutic,Marra,Zutic2,Zutic3,FPTA} and experimental~\cite{Kontos,FrolovMagnets} attention in the field of engineered topological superconductivity. Furthermore, in a different context, recent experiments~\cite{CoFe2C} have found evidence for colossal magnetic anisotropy for CoFe$_2$C magnets with a cha\-ra\-cte\-ri\-stic dimension of the order of few nanometers. Since the wave vectors controlling the spatial periodicity of the MTC are required to roughly match those connecting pairs of nodes, it may be currently challenging for these state-of-the art nanomagnetic technologies to be applicable in most of the above listed SCs. This is because most of these are me\-tals and thus the arising nodes are expected to be connected by wave vectors with a length which relates to the Fermi wave number. Hence, we conclude that the present route for engineering vortex MZMs, by means of impo\-sing MTCs induced by magnets, appears more relevant for very low-density and bad-metal SCs, with Fermi wave numbers in the few nanometer regime.

\subsection{Spontaneously Induced MTCs}

Another possibility is the interaction-driven MTCs, which can become stabilized in the presence of attractive interactions in the magnetic channel, in order to minimize the free energy of the system. In analogy to 1D spin-density waves~\cite{Gruner}, which are promoted by the perfect ne\-sting of the two points comprising the Fermi surface, here we expect a MTC to spontaneously appear and gap out the nodes. This is under the condition that  other competing instabilities are subdominant to the MTC. The tendency of the system towards the spontaneous development of a magnetic helix cry\-stal which gaps out a single pair of nodes can be here inferred by eva\-lua\-ting the respective spin susceptibility. In fact, the latter can be calculated using the low-energy model of Eq.~\eqref{eq:NodeHamiltonianProjected}. 

For this purpose, we restrict to the $n$-th pair of nodes and consider a magnetic helix crystal with projected components $\bm{M}_\zeta^{(n)}=\big(M_{\zeta;x}^{(n)},M_{\zeta;y}^{(n)},0\big)$. Moreover, we assume that the magnetic helix crystal is governed by a pe\-rio\-di\-ci\-ty given by a wave vector which is equal to the wave vector $2\bm{k}_n$ connecting the nodes of the pair of interest. We note that, when the wave vector of the magnetic helix matches the one connecting the nodes, a gap opens on these with an infinitesimally weak strength of $|\bm{M}_\zeta^{(n)}|$. Nonetheless, a full gap is accessible also for detuned wave vectors, but in that event, a threshold strength of $|\bm{M}_\zeta^{(n)}|$ has to be reached. In the following, we restrict to the ideal scenario of perfectly matched wave vectors, since this is the configuration that yields the highest susceptibility and thus minimizes the free energy of the system.

In the upcoming analysis, we suppress the $^{(n)}$ index from the various variables to simplify the notation. Since in the absence of magnetism the Hamiltonian in Eq.~\eqref{eq:NodeHamiltonianProjected} is dictated by an emergent U(1) symmetry generated by $\lambda_z$, we can evalua\-te the desired susceptibility by con\-si\-de\-ring that only one of the components of the magnetic helix texture, either $M_{\zeta,x}$ or $M_{\zeta,y}$, is nonzero. Under the above conditions, straightforward calculations in the zero temperature limit yield the susceptibility expression:
\bea
\chi=\int\frac{d\bm{q}}{(2\pi)^2}\frac{1}{\sqrt{\left[\bm{q}\cdot\bm{\upsilon}_{\varepsilon,\zeta}\right]^2+\left[\bm{q}\cdot\bm{\upsilon}_{\Delta,\zeta}\right]^2}}\,.
\eea

\noi By transferring to polar coordinates $\bm{q}=q\big(\cos\gamma,\sin\gamma\big)$, and after introducing a cutoff wavenumber $q_c$, we end up with the expression:
\begin{align}
\chi=\int_0^{2\pi}\frac{d\gamma}{2\pi}
\frac{\nu_c}{\sqrt{1+\cos\gamma_0\cos\big(2\gamma\big)+\delta\sin\gamma_0\sin\big(2\gamma\big)}}\label{eq:SpontanSusc}
\end{align}

\noi where we introduced the density of states $\nu_c=q_c/\big(2\pi\bar{\upsilon}\big)$ which depends on the cutoff $q_c$ and the average velocity defined as $\bar{\upsilon}=\sqrt{\big(|\bm{\upsilon}_{\varepsilon,\zeta}|^2+|\bm{\upsilon}_{\Delta,\zeta}|^2\big)/2}$. The final outcome for $\chi$ is decided by the precise values of the parameter $\delta=\big(|\bm{\upsilon}_{\varepsilon,\zeta}|^2-|\bm{\upsilon}_{\Delta,\zeta}|^2\big)/\big(|\bm{\upsilon}_{\varepsilon,\zeta}|^2+|\bm{\upsilon}_{\Delta,\zeta}|^2\big)$ which encodes the velocity mismatch, and the relative angle $\gamma_0=\gamma_\varepsilon-\gamma_\Delta$, which is defined in terms of the orientation angles $\gamma_{\varepsilon,\Delta}$ given by $\bm{\upsilon}_{\varepsilon,\zeta}=|\bm{\upsilon}_{\varepsilon,\zeta}|\big(\cos\gamma_\varepsilon,\sin\gamma_\varepsilon\big)$ and $\bm{\upsilon}_{\Delta,\zeta}=|\bm{\upsilon}_{\Delta,\zeta}|\big(\cos\gamma_\Delta,\sin\gamma_\Delta\big)$. We note that the final form of Eq.~\eqref{eq:SpontanSusc} was obtained after the redefinition $\gamma\mapsto\gamma+(\gamma_\varepsilon+\gamma_\Delta)/2$, which does not influence the result.

From Eq.~\eqref{eq:SpontanSusc} we infer that in contrast to 1D spin-density waves, here, the 2D dimensionality does not ge\-ne\-ral\-ly allow for a logarithmically divergent suscepti\-bi\-li\-ty. In fact, for velocity vectors oriented at right angles ($\gamma_0=\pi/2$) which additionally feature equal lengths ($\delta=0$), the susceptibility reaches its minimum value $\chi_0=\nu_c$. Away from this highly symmetric configuration, the ratio $\chi/\chi_0$ increases monotonically. Notably, the su\-sce\-ptibility becomes maximized as the system tends to the extreme anisotropic cases with $\gamma_0/\pi\rightarrow\mathbb{Z}$ for arbitrary $\delta$, or, $|\delta|\rightarrow1$ for arbitrary $\gamma_0$. Thus, the appea\-ran\-ce of a MTC generally requires a threshold strength for the interaction which drives magnetism. Noteworthy, this cri\-ti\-cal strength depends strongly on the cutoff $q_c$, since $\chi$ scales linearly with $q_c$. Finally, this threshold interaction strength becomes reduced by enhancing the anisotropy of the energy di\-spersion in the vicinity of the nodes. 

Out of the various candidates mentioned earlier, Fe-SCs can support nodal phases, exhibit single- and double-$\bm{Q}$ magnetic stripe order~\cite{avci14a,wasser15,hassinger,bohmer15a,allred15a,zheng16a,malletta,mallettb,pratt,allred16a,meier17}, and can harbor the microscopic coexi\-sten\-ce of magnetism and supercon\-duc\-ti\-vi\-ty~\cite{avci14a,wang16a,johrendt11,avci11,klauss15,ni08a,nandi10a}. Moreover, recent theo\-re\-ti\-cal studies~\cite{Christensen_18} predict single- and double-$\bm{Q}$ MTCs in doped 122 and 1111 compounds. While the theoretical investigation and experimental support regarding the microscopic coe\-xi\-sten\-ce of nodal Fe-SCs and MTCs is still lacking, these sy\-stems appear as potential candidates to observe the phenomenology discussed in this section.

\subsection{MTCs Induced by Localized Magnetic Moments}

The last possible physical realization of our theory relies on nodal SCs coupled with lattices of lo\-ca\-li\-zed magnetic moments. Here, the MTC required for gapping out the nodes of the SC is considered to result from the magnetization of the localized moments. This third mechanism that we propose for engineering MZMs bears similarities to the ones proposed~\cite{NadgPerge,Nakosai,Braunecker,Klinovaja,Vazifeh,Pientka,Ojanen,Sedlmayr,Paaske1,Paaske2,Brydon,Li,Heimes2,PKSTM} for MZM platforms relying on magnetic chains~\cite{Yazdani1,Ruby,Meyer,Yazdani2,Gerbold,Wiesendanger,Cren,GerboldEUphys}. However, here, the dimension of the lattice of the magnetic impurities needs to coincide with the dimension of the lattice defined by the nodal SC, akin to the picture that holds for Kondo lattice systems~\cite{Doniach}, which are typical scenarios for rare-earth~\cite{Ivar} and heavy-fermion SCs~\cite{BrydonSchnyder,Sigrist}. Moreover, we remind the reader that within our proposal MZMs are trapped by vortices of the MTCs instead of termination edges or domain walls. Interestingly, employing MTC vortices for trapping MZMs appears parti\-cu\-lar\-ly attractive in the case of hybrid sy\-stems invol\-ving magnetic adatoms, since MTC vortices can be in principle tailored using spin-polarized scanning tunneling microscopy~\cite{Yazdani2,Wiesendanger,Cren,GerboldEUphys}. The latter technique, further allows for the spin-sensitive detection of the vortex MZMs~\cite{PKSTM}.

In a similar fashion to proposals for self-organized MZM platforms using magnetic adatom chains~\cite{Braunecker,Klinovaja,Vazifeh}, also here, we expect for the magnetization of the magnetic moment lattice to exhibit a spatial profile which reflects the structure of the nodes of the host SC. This profile is determined by the spin susceptibility $\chi_{nm}^{\alpha\beta}$ which couples the magnetic moments through the Ruderman-Kittel-Kasuya-Yosida (RKKY) type of energy term:
\bea
E_{\rm RKKY}=-\frac{J^2}{2}\sum_{n,m}\sum_{\alpha,\beta}S_n^\alpha\chi_{\alpha\beta}(\bm{R}_n-\bm{R}_m)S_m^\beta\,,
\eea

\noi where the indices $n,m$ label the impurity lattices sites $\bm{R}_{n,m}$, while $\alpha,\beta=x,y,z$ label the spin components of the moments. $J$ denotes the strength of the magnetic exchange coupling between the moment and the electron spin. For a homogeneous nodal SC of a spatial dimension $d$, the spin susceptibility depends only on the dif\-fe\-ren\-ce $\bm{R}=\bm{R}_n-\bm{R}_m$ of the position vectors of the coupled magnetic moments and is defined by the expression:
\begin{align}
\chi_{\alpha\beta}(\bm{R})=-\frac{1}{2}\int_{-\infty}^{+\infty}\frac{d\epsilon}{2\pi}{\rm Tr}\big[\sigma_\alpha\hat{G}_0(\epsilon,\bm{R})\sigma_\beta\hat{G}_0(\epsilon,-\bm{R})\big]
\label{eq:SpinSuscAdatom}
\end{align}

\noi where we introduced the bare matrix Green function:
\begin{align}
\hat{G}_0(\epsilon,\bm{R})=\int\frac{d\bm{k}}{(2\pi)^d}\ph e^{i\bm{k}\cdot\bm{R}}\ph\hat{G}_0(\epsilon,\bm{k})\,,
\end{align}

\noi with its momentum space counterpart being defined through: $\hat{G}_0^{-1}(\epsilon,\bm{k})=i\epsilon-\hat{\cal H}_0(\bm{k})$, where $\hat{\cal H}_0(\bm{k})$ is given by Eq.~\eqref{eq:H0}. We introduced a factor of $\nicefrac{1}{2}$ in Eq.~\eqref{eq:SpinSuscAdatom} to prevent the double counting of the electronic degrees of freedom since we use the four-component basis of Eq.~\eqref{eq:Spinor}.

To further elaborate on this aspect, we explore a concrete example for a nodal SC with two pairs of nodes. Specifically, we consider the $d=2$ model with $\upsilon_\Delta>0$:
\begin{align}
\hat{\cal H}_0(\hat{\bm{p}})=\tau_z\bigg(\frac{\hat{p}_y^2}{2m}-\alpha\hat{p}_y\sigma_z-\mu\bigg)+\upsilon_\Delta \hat{p}_x\tau_x\sigma_z\label{eq:PrototypicalModel}
\end{align}

\noi which is obtained by means of dimensional reduction of the 3D model described in Eq.~\eqref{eq:3Dmodel} to the 2D $(x,y)$ plane, and harbors two pairs of nodal points located along the $k_x=0$ high-symmetry line. The above model constitutes a paradigmatic model for the entire category of systems discussed here, since all nodes are situated along the same axis, thus allowing the generalization of some of our results to nodal SCs harboring multiple pair of nodes, with the various pairs of nodes been situated on lines with generally non-parallel orientation. 

Before carrying out the calculation of the spin-susceptibility, it is convenient to gauge away the SOC term by performing a spatially-dependent unitary transformation:
\begin{align}
\hat{\cal H}_0(\hat{\bm{p}})={\cal O}(y)\left(\frac{\hat{p}_y^2-k_F^2}{2m}\tau_z+\upsilon_\Delta \hat{p}_x\tau_x\sigma_z\right){\cal O}^\dag(y)\,,\label{eq:PrototypicalModelUnitary}
\end{align}

\noi with $k_F^2/(2m)=\mu+m\alpha^2/2$ and the kinetic energy term $\varepsilon(k_y)=(k_y^2-k_F^2)/(2m)$. Moreover, the unitary transformation matrix reads as: ${\cal O}(y)={\rm Exp}\left(im\alpha y\sigma_z\right)$. Within this framework, the bare Green function in coordinate space becomes $\hat{G}_0(\epsilon,\bm{R})={\cal O}(Y)\hat{G}_0^{\alpha=0}(\epsilon,\bm{R})$ where we introduced $\bm{R}=(X,Y)$, and the Green function for $\alpha=0$:
\begin{align}
\hat{G}_0^{\alpha=0}(\epsilon,\bm{R})=\int\frac{d\bm{k}}{(2\pi)^2}\ph e^{i\bm{k}\cdot\bm{R}}\ph\frac{i\epsilon+\varepsilon(k_y)\tau_z+\upsilon_\Delta k_x\tau_x\sigma_z}{(i\epsilon)^2-\big(\upsilon_\Delta k_x\big)^2-[\varepsilon(k_y)]^2}.
\end{align}

We now proceed with the evaluation of the spin susceptibility. For this purpose, we also consider a semiclassical approach in which $k_F$ is substantial, thus al\-lo\-wing to approximate the kinetic energy as $\varepsilon(k_y)\approx\pm\upsilon_F\big(k_y\mp k_F\big)$, with the latter being expanded about the two Fermi points $k_y=\pm k_F$ with $\upsilon_F=k_F/m$. Moreover, due to the presence of the strong anisotropy, it is preferrable to integrate over the $k_x$ variable first. Since the integrand peaks at $k_x=0$, we can safely consider that the integration is over $(-\infty,+\infty)$, without wor\-rying about the possible necessity to introduce a cutoff. In this case, employing standard residue theory provides:
\bea
\hat{G}_0^{\alpha=0}(\epsilon,\bm{R})&=&-\int\frac{dk_y}{4\pi\upsilon_\Delta}\ph e^{ik_yY-\sqrt{[\varepsilon(k_y)]^2+\epsilon^2}|X|/\upsilon_\Delta}\bigg\{
\no\\
&&\frac{i\epsilon+\varepsilon(k_y)\tau_z}{\sqrt{[\varepsilon(k_y)]^2+\epsilon^2}}+i{\rm sgn}{\big(X\big)}\tau_x\sigma_z\bigg\}\,.
\eea

\noi By now exploiting the assumed semiclassical limit, we introduce the variable $\xi=\pm\upsilon_F(k_y\mp k_F)$ which allows us to carry out the substitution $k_y=\pm(k_F+\xi/\upsilon_F)$. After taking the limit $k_F\rightarrow\infty$, we obtain for $X,Y\neq0$:
\bea
&&\hat{G}_0^{\alpha=0}(\epsilon,\bm{R})
=\frac{\cos\big(k_FY\big)}{\upsilon_F\upsilon_\Delta}\int_{-\infty}^{+\infty}\frac{d\xi}{2\pi}e^{i\xi Y/\upsilon_F-\sqrt{\xi^2+\epsilon^2}|X|/\upsilon_\Delta}\bigg\{\no\\
&&\qquad\qquad\quad-\frac{i\epsilon}{\sqrt{\xi^2+\epsilon^2}}-i{\rm sgn}{\big(X\big)}\tau_x\sigma_z\bigg\}\no\\
&&\phd
-\frac{\sin\big(k_FY\big)}{\upsilon_\Delta}\frac{d}{dY}\int_{-\infty}^{+\infty}\frac{d\xi}{2\pi}\ph\frac{e^{i\xi Y/\upsilon_F-\sqrt{\xi^2+\epsilon^2}|X|/\upsilon_\Delta}}{\sqrt{\xi^2+\epsilon^2}}\tau_z\,.
\eea

\noi By further restricting to situations where $|X|$ is small with $\Theta=|X/Y|\ll1$, we manage to obtain appro\-xi\-ma\-te closed-form expressions by replacing the exponential term ${\rm Exp}\big(-\sqrt{\xi^2+\epsilon^2}|X|/\upsilon_\Delta)$ by its factorized form ${\rm Exp}(-|\xi||X|/\upsilon_\Delta){\rm Exp}(-|\epsilon||X|/\upsilon_\Delta)$ when evaluating the term $\propto\tau_x\sigma_z$, and by completely discar\-ding it when eva\-lua\-ting the remaining two terms. These approximations lead to the expression:
\bea
&&\pi\upsilon_F\upsilon_\Delta\hat{G}_0^{\alpha=0}(\epsilon,\bm{R})
\approx\sin\big(k_F|Y|\big)|\epsilon|K_1\big(|\epsilon||Y|/\upsilon_F\big)\tau_z\no\\
&&-\cos\big(k_FY\big)
\left[i\epsilon K_0\big(|\epsilon||Y|/\upsilon_F\big)+\frac{i\upsilon_F^2Xe^{-|\epsilon||X|/\upsilon_\Delta}}{\upsilon_\Delta Y^2}\tau_x\sigma_z\right].\no\\
\eea

With the Green function at hand, we now determine the elements of the spin-susceptibility tensor. We find that for $\alpha=0$, the only nonzero elements are $\chi_{xx}^{\alpha=0}(\bm{R})=\chi_{yy}^{\alpha=0}(\bm{R})\equiv\chi_{\rm in}^{\alpha=0}(\bm{R})={\cal I}_1(\bm{R})-{\cal I}_2(\bm{R})+{\cal I}_3(\bm{R})$ and $\chi_{zz}^{\alpha=0}(\bm{R})={\cal I}_1(\bm{R})-{\cal I}_2(\bm{R})-{\cal I}_3(\bm{R})$, which consist of the three non-negative contributions presented below:
\bea
{\cal I}_1(\bm{R})&=&\frac{1+\cos\big(2k_FY\big)}{2\pi\big(\pi\upsilon_\Delta Y\big)^2}\frac{\upsilon_F}{|Y|}\int_{-\infty}^{+\infty}du\ph u^2K_0^2(|u|)\,,\quad\\
{\cal I}_2(\bm{R})&=&\frac{1-\cos\big(2k_FY\big)}{2\pi\big(\pi\upsilon_\Delta Y\big)^2}\frac{\upsilon_F}{|Y|}\int_{-\infty}^{+\infty}du\ph u^2K_1^2(|u|)\,,\\
{\cal I}_3(\bm{R})&=&\frac{\cos^2\big(k_FY\big)}{\pi\big(\pi\upsilon_\Delta Y\big)^2}\frac{\upsilon_F}{|Y|}\frac{\upsilon_F}{\upsilon_\Delta}\Theta\,.
\eea

\noi After the evaluation of $\int_{-\infty}^{+\infty}du\ph u^2K_0^2(|u|)=\pi^2/16$ and $\int_{-\infty}^{+\infty}du\ph u^2K_1^2(|u|)=3\pi^2/16$, we find the following result:
\begin{align}
\chi_{{\rm in},zz}^{\alpha=0}(\bm{R})\approx\frac{2\cos\big(2k_FY\big)-1\pm\Theta\frac{\upsilon_F}{\upsilon_\Delta}\big[\frac{4}{\pi}\cos\big(2k_FY\big)\big]^2}{16\pi\big(\upsilon_\Delta Y\big)^2|Y|/\upsilon_F}.
\end{align}

From the above, we conclude that when $\alpha=0$ and two magnetic moments are placed along a line which is pa\-ral\-lel to the $y$ direction, the RKKY coupling is spin-isotropic since $\Theta=0$. This is to be expected since, in this case, the $k_x=0$ momentum is involved in the calculation of the Green function, which coincides with the line, along which, the pairs of nodes appear for the Hamiltonian of Eq.~\eqref{eq:PrototypicalModel}. From the $Y$ dependence of the spin-susceptibility, we find that both ferromagnetic (FM) and antiferromagnetic (AFM) solutions are accessible. In contrast, when the vector con\-nec\-ting the positions of two magnetic moments forms an angle with the $y$ axis, then an easy plane/axis spin anisotropy sets in, and can be traced back to the spin-dependent p-wave pairing term in Eq.~\eqref{eq:PrototypicalModel}. For $\Theta\ll1$, we find that when the distance between two moments is such so that it favors FM orde\-ring, the spin-anisotropy tends to align the magnetic moments in the plane, since in this case $\chi_{\rm in}^{\alpha=0}>\chi_{zz}^{\alpha=0}$. On the other hand, for an inter-moment distance favoring AFM orde\-ring, the spin-anisotropy forces the magnetic moment to point out-of-the plane, i.e., along the $z$ axis. 

Let us now consider the effects of nonzero $\alpha$ which implies that the Rashba SOC in non-negligible. For a perfectly ordered array of localized magnetic moments with a spacing $|Y|$ in the $y$ direction, we find that the $z$ axis susceptibility is unaltered and the components $\chi_{xz,zy}$ remain zero. Nonetheless, there is now a modification of the strengths of the in plane suscepti\-bi\-li\-ties. In addition to the above, the inplane off diagonal ele\-ments $\chi_{xy}(\bm{R})=-\chi_{yx}(\bm{R})$ become now nonzero. These results are reflected in the expressions:  
\bea
\chi_{\rm in}(\bm{R})&=&\cos\big(2m\alpha Y\big)\chi_{\rm in}^{\alpha=0}(\bm{R})\,,\\
\chi_{xy}(\bm{R})&=&\sin\big(2m\alpha Y\big)\chi_{\rm in}^{\alpha=0}(\bm{R})\,.
\eea

\noi Notably, a similar susceptibility structure has been di\-scussed previously in Ref.~\onlinecite{Heimes2} in connection to MZM platforms with magnetic chains. Here, the obtained spin-susceptibility transforms the emergent in plane FM order into an inplane magnetic helix texture, while it leaves the out-plane AFM order unaffected. Since the $z$ components of the magnetization become projected out in the low-energy model of Eq.~\eqref{eq:NodeHamiltonianProjected} only the inplane ordering of the magnetic moments can gap out the nodes. However, in the absence of additional sources of spin anisotropy, the arising FM-like magnetic helix texture cannot gap out the outer-helical branch nodes of the model in Eq.~\eqref{eq:PrototypicalModel}, since its winding compensates that of the Rashba SOC. A possible escapeway is to impose or engineer additional inplane spin-anisotropies, which can allow to the magnetic moments to either develop inplane magnetic helix textures with a different pitch than the one cha\-rac\-te\-ri\-zing the Rashba SOC, or enable in plane AFM phases. 

For example the addition of terms such as $\big(S_n^x\big)^2$ ge\-ne\-ral\-ly opens the door for the above scenarios, as it has been already demonstrated in Ref.~\onlinecite{Heimes2}. So far, we have not made any mentioning in connection to the pair of nodes associated with the inner helical branch arising due to the Rashba SOC. In analogy to Sec.~\ref{sec:SectionVIII}, also here, we can assume the presence of a weak Zeeman field in order to lift the inner helical branch~\cite{LutchynPRL,OregPRL}. Finally, we wish to remind the reader that the present platforms appear attractive for pinning MZMs in vortices of the arising inplane magnetic helix texture or AFM order, since the spin-orientation of the magnetic moments is in principle locally tunable using scanning tunneling microscopy. 

\section{Summary and Outlook}\label{sec:SectionX}

We provide a novel pathway to engineer Majorana zero modes (MZMs), which relies on nodal superconductors (SCs) and at the same time does not require the pre\-sen\-ce of superconducting vortices. In contrast, we show that MZMs become accessible in nodal SCs which are under the influence of magnetic texture crystals (MTCs). At this point, we wish to clarify that our approach is distinct to Refs.~\onlinecite{Cren,NakosaiBalatsky,KlinovajaSkyrmion,Mirlin,Garnier,Mesaros}. There, isolated magnetic skyrmions, and not cry\-stals as we consider here, have been employed as smooth defects which pin various types of bound states in fully-gapped SCs, these inclu\-ding MZMs~\cite{Cren,KlinovajaSkyrmion,Mirlin,Garnier}. Instead, within our proposal, MZMs are pinned by spin and/or shift vortices induced in the MTC, which constitute singular defects. 

Our analysis provides a detailed study of va\-rious topological scenarios in both 2D and 3D nodal SCs, which cover all three Majorana symmetry classes, i.e., BDI, D, and DIII. We present in detail the construction for two topological invariant quantities, which predict vortex MZMs in class BDI in 2D and chiral Majorana modes in class D in 3D. Topological invariants for the remaining cases can be constructed using these two fundamental invariants. Even more, we show how to render Majorana Kramers pair solutions still accessible, in spite of the vio\-lation of the standard time-reversal symmetry by the MTC. As an example we discuss a two-band model in 2D which harbors a single vortex Kramers pair of MZMs in 2D. Our analytical approach and predictions based on the various topological invariants are backed by our numerical simulations on the lattice. 

The last part of our investigation concerns the expe\-ri\-men\-tal realization of our proposals in connection to intrinsic nodal SCs. We discuss the following three different possible paths to engineer the required MTCs and concomitant vortex MZMs: (i) MTCs which are engineered using nanomagnets, (ii) MTCs that arise spontaneously by virtue of the interactions dictating the electrons of the SC, and (iii) MTCs which are harbored by a lattice of localized magnetic moments which are exchange-coupled to the nodal SC. We provide insight in the above possibilities and discuss their advantages and possible limitations. Regarding the stabilization of vortices in MTCs, we note that certain types of to\-po\-lo\-gi\-cal defects in MTCs, e.g., disclinations~\cite{Nattermann,NattermannRev}, have already been experimentally observed in helimagnets~\cite{UchidaFeGe,MagneticMonopole,TopoDWHelimagnets1,DefectsMnSi,TopoDWHelimagnets2}. Similarly, shift/spin vortices can arise spontaneously, get pinned by disorder or engineered with nanomagnets, or scanning tunneling microscopy (STM) tips.

Apart from quantum materials, our idea may also be relevant for artificial platforms, such as a 2D electron gas (2DEG) in pro\-xi\-mi\-ty to a conventional SC with strong Rashba SOC. Here, one wishes to harness the pro\-xi\-mi\-ty effect to induce a mixed-spin type nodal pairing term in the 2DEG. This appears particularly promising for superconductor-semiconductor devices for which a strong debate has been raised recently~\cite{PradaRev,CXLiu,Moore,Reeg,Woods,FrolovUbiquitous,Vuik,PanDasSarma,YuQuasiMZM}. We hope that our theory inspires the development of new platforms which take advantage of the advanced fabrication and measurement techniques which currently exist for semiconductor hybrids, and combine them with the control over MTCs and vortex MZMs using spin-polarized STM.

\section*{Acknowledgements}

We thank M.~H. Christensen and A.~V. Ba\-latsky for inspiring discussions. D.~S. and B.~M.~A. acknowledge support from the Carlsberg Foundation, and P.~K. from the National Natural Science Foundation of China (Grant No. 12050410262). B.~M.~A. additionally acknowledges support from the Independent Research Fund Denmark grant number 8021-00047B.

\end{document}